\documentclass[11pt,onecolumn]{IEEEtran}
\usepackage{amsthm,amsmath}
\usepackage{epsfig,amssymb,amsbsy,verbatim,subfigure,array}
\usepackage{pstricks,cite,url}%caption
\usepackage{multicol,afterpage,paralist} %fancyheadings,
\usepackage{hyperref}

\def\l{\ell}

\def\detF{\operatorname{detf}}

\def\calP{\mathcal{P}}
\def\calR{\mathcal{R}}

\def\L{\mathcal{L}}

\def\ie{\emph{i.e.}}
\def\eg{\emph{e.g.}}
\def\E{\mathbb{E}}
\def\P{\mathbb{P}}
\def\R{\mathbb{R}}
\def\N{\mathbb{N}}

\def\figref#1{Fig.~\ref{#1}}

\def\argmin{\operatorname{arg~min}}

\def\i{\mathbf{1}}
\def\sinc{\operatorname{sinc}}

\def\misr{\mathsf{MISR}}
\def\efir{\mathsf{EFIR}}
\def\SIR{\mathsf{SIR}}
\def\sir{\SIR}
\def\isr{\mathsf{I{\bar S}R}}

\def\one{\mathbf{1}}

\def\dd{{\rm d}}

\def\ps{p_{\rm s}}
\def\tps{\tilde{p}_{\rm s}}
\def\psref{p_{\rm s,ref}}
\def\psppp{p_{\rm s,PPP}}
\def\T{\theta}
\def\calG{\mathcal{G}}
\def\calL{\mathcal{L}}

\newtheorem{theorem}{Theorem}

\newtheorem{lemma}{Lemma}
\newtheorem{corollary}[theorem]{Corollary}%[theorem]
\newtheorem{definition}{Definition}

\def\mh#1{}%{{\blue\sffamily\small\em $\Rightarrow$ #1 $\Leftarrow$}}
\def\rk#1{}%{{\red\sffamily\small\em $\Rightarrow$ #1 $\Leftarrow$}}

\let\arxiv\iftrue%\iffalse%\iffalse%\iftrue

\newlength{\figwidth}
\begin{document}
\title{Asymptotics and Approximation of the SIR\\Distribution in General Cellular Networks}
\author{Radha Krishna Ganti, \IEEEmembership{Member, IEEE}, and Martin Haenggi, \IEEEmembership{Fellow, IEEE}
\thanks{Manuscript date \today. Part of this work was presented at the 2015 IEEE International Symposium on Information
Theory (ISIT'15) \cite{net:Ganti15isit}. Radha Krishna Ganti is with the Indian Institute of Technology, Madras, India (e-mail: \url{rganti@ee.iitm.ac.in}), and Martin Haenggi is with the University of Notre Dame, Notre Dame, IN, USA (e-mail: \url{mhaenggi@nd.edu}).
The support of the NSF (grants CCF 1216407 and CCF 1525904) is gratefully acknowledged.}
}
\maketitle
\begin{abstract}
It has recently been observed that the SIR distributions
of a variety of cellular network models and transmission techniques
look very similar in shape. As a result, they are well approximated
by a simple horizontal shift (or gain) of the distribution of the most tractable
model, the Poisson point process (PPP). To study and explain this
behavior, this paper focuses on general single-tier network models with nearest-base
station association and studies the asymptotic gain both at 0 and at infinity.

We show that the gain at 0 is determined
by the so-called mean interference-to-signal ratio (MISR) between the 
PPP and the network model under consideration, while the gain at infinity
is determined by the expected fading-to-interference ratio (EFIR).

The analysis of the MISR is based on a novel type of point process, the so-called relative distance process,
which is a one-dimensional point process on the unit interval [0,1] that fully determines the SIR.
A comparison of the gains at 0 and infinity shows that
the gain at 0 indeed provides an excellent approximation for the
entire SIR distribution. Moreover, the gain is mostly a function of the network geometry
and barely depends on the path loss exponent and the fading. 
The results are illustrated using several examples of repulsive point processes.
\end{abstract}
\begin{IEEEkeywords}
Cellular networks, stochastic geometry, signal-to-interference ratio, Poisson point processes.
\end{IEEEkeywords}

\section{Introduction}
\subsection{Motivation}
The distribution of the signal-to-interference ratio (SIR) is a key quantity in the analysis
and design of interference-limited wireless systems. Here we focus on general single-tier
cellular networks where users are connected to the strongest (nearest)
base station (BS). Let $\Phi\subset\R^2$ be a point process
representing the locations of the BSs and
let $x_0\in\Phi$ be the serving BS of the typical user at the origin, \ie,
define $x_0\triangleq\argmin\{x\in\Phi\colon \|x\|\}$. Assuming all BSs transmit at the same power
level, the downlink SIR is given by
\begin{equation}
   \sir\triangleq \frac{S}{I}=\frac{h_{x_0} \ell(x_0)}{\sum_{x\in\Phi\setminus\{x_0\}} h_x \ell(x)}, 
  \label{sir}
\end{equation}
where $(h_x)$ are iid random variables representing the fading and $\ell$ is the
path loss law.
The complementary cumulative distribution (ccdf) of the SIR is
\begin{equation}
\bar F_{\sir}(\theta)\triangleq \P(\sir>\theta).
\end{equation}
Under the SIR threshold model for reception, the ccdf of the SIR can also be interpreted
as the success probability of a transmission, \ie, $\ps(\theta)\equiv \bar F_\sir(\theta)$.

In the case where $\Phi$ is a homogeneous Poisson point process (PPP), Rayleigh fading,
 and $\ell(x)=\|x\|^{-\alpha}$, the success probability was determined in
\cite{net:Andrews11tcom}. It can be expressed in terms of the 
Gaussian hypergeometric function $_2F_1$
as \cite{net:Zhang14twc}
\begin{equation}
   \psppp(\theta)=\frac{1}{_2F_1(1,-\delta; 1-\delta; -\theta)} ,
  \label{ps_ppp}
 \end{equation}
where $\delta\triangleq 2/\alpha$. For $\alpha=4$, remarkably, this simplifies to
\[ \psppp(\theta)=\frac1{1+\sqrt\theta\arctan\sqrt\theta} .\]
In \cite{net:Nigam14tcom}, it is shown that the same expression holds for the homogeneous
independent Poisson (HIP) model, where the different tiers in a heterogeneous
cellular network form independent homogeneous PPPs.
For all other cases, the success probability is intractable or can at best be
expressed using combinations of infinite sums and integrals. Hence there is
a critical need for techniques that yield good approximations of the SIR
distribution for non-Poisson networks.

\subsection{Asymptotic SIR gains and the MISR}
It has recently been observed in \cite{net:Guo15tcom,net:Haenggi14wcl}
that the SIR ccdfs for different point processes
and transmission techniques (e.g., BS cooperation or silencing) {\em appear to be merely horizontally shifted
versions of each other} (in dB), as long as their diversity gain is the same.

Consequently, the success probability of a network model can be accurately approximated by that
of a reference network model by scaling the threshold $\theta$ by this SIR gain factor (or shift in dB)
$G$, \ie,
\[  \ps(\theta) \approx \psref(\theta/G) .\]

Formally, the horizontal gap at target probability $p$ is defined as
\begin{equation}
G_{\rm p}(p)\triangleq\frac{\bar F_{\sir}^{-1}(p)}{\bar F_{\sir_{\rm ref}}^{-1}(p)},\quad p\in(0,1),
\label{gp}
\end{equation}
where $\bar F_{\sir}^{-1}$ is the inverse of the ccdf of the SIR and $p$ is the
success probability where the gap is measured.
It is often convenient to consider the gap as a function of $\theta$, defined as
\begin{equation}
 G(\theta)\triangleq G_{\rm p}(\psref(\theta))=\frac{\bar F_{\sir}^{-1}(\psref(\theta))}{\theta} .
 \label{g_theta}
\end{equation}
Due to its tractability, the PPP is a sensible choice
as the reference model\footnote{This is why the method of approximating an SIR distribution
by a shifted version of the PPP's SIR distribution is called ASAPPP---``Approximate SIR
analysis based on the PPP" \cite{net:Haenggi14hetsnets}.}.

If the shift $G(\theta)$ is indeed approximately a constant, \ie, $G(\theta)\approx G$, then
$G$ can be determined by evaluating $G(\theta)$ for an arbitrary value of $\theta$. As shown
in \cite{net:Haenggi14wcl}, the limit of $G(\theta)$ as $\theta\to 0$ is relatively easy to calculate. Here
we focus in addition on the positive limit $\theta\to\infty$ and compare the two asymptotic
gains to demonstrate the effectiveness of the idea of horizontally shifting SIR distributions 
by a constant.
 
So the main focus of this paper are the asymptotic gains relative
to the PPP, defined as follows.
\begin{definition}[Asymptotic gains relative to PPP]
The asymptotic gains (whenever the limits exist) $G_0$ and $G_\infty$ are defined as
\begin{equation}
G_0\triangleq \lim_{\theta\to 0} G(\theta); \quad G_\infty\triangleq \lim_{\theta\to \infty} G(\theta),
\label{gp2}
\end{equation}
where the PPP is used as the reference model.
\end{definition}
\subsection{Prior work}
Some insights on $G_0$ are available from prior work. In  \cite{net:Haenggi14wcl} it is shown
that for Rayleigh fading, $G_0$ is closely connected to the mean interference-to-signal ratio (MISR).
The MISR is the mean of the interference-to-(average)-signal ratio ISR, defined as
\[ \isr\triangleq \frac{I}{\E_h(S)} , \]
where $\E_h(S)=\E(S\mid\Phi)$ is the mean received signal power averaged only
over the fading. 
Not unexpectedly, the calculation of the MISR for the PPP is relatively straightforward and yields 
$\misr_{\rm PPP}=2/(\alpha-2)$ \cite[Eqn.~(8)]{net:Haenggi14wcl}\footnote{A different
derivation of this result will be given in Thm.~\ref{thm:misr_pp} in Sec.~IV in this paper.}.
Since $\P(\sir>\theta)=\P(h>\theta\,\isr)=\E(\P(h>\theta\,\isr\mid\isr))$,
the success probability can in general be expressed as
\begin{equation}
\ps(\theta)=\E\bar F_h(\theta\,\isr) ,
\label{ps_misr}
\end{equation}
where $\bar F_h$ is the ccdf of the fading random variables.
For Rayleigh fading, $\bar F_h(x)=e^{-x}$ and thus $\ps(\theta)\sim 1-\theta\,\misr$,
$\theta\to 0$, resulting in 
\[ G_0= \frac{\misr_{\rm PPP}}{\misr}= \frac{2}{\alpha-2}\:\frac1{\misr} \]
 and
\[ \ps(\theta) \sim \psppp(\theta/G_0), \quad\theta\to 0.\]
So, asymptotically, shifting the ccdf of the SIR distribution of the PPP is exact. 

\begin{figure}
\centerline{\includegraphics[width=\figwidth]{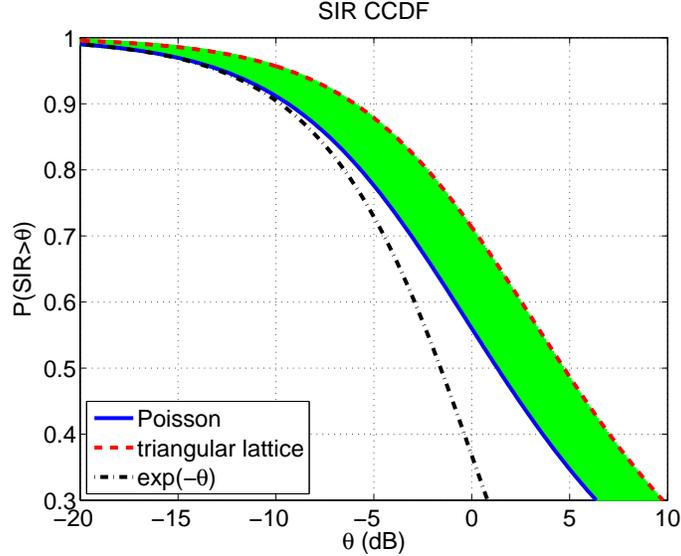}}
\caption{The SIR distributions for the PPP (solid) and the triangular lattice (dashed) for $\alpha=4$
and the lower bound (which is asymptotically tight) $e^{-\theta}$ for the PPP (dash-dotted).
The horizontal gap between the SIR distributions of the PPP and the triangular lattice is 3.4 dB
for a wide range of $\theta$ values. The shaded band indicates the region in which the SIR distributions for
all stationary point process fall that are more regular than the PPP.}
\label{fig:sir_gap}
\end{figure}

An example is
shown in \figref{fig:sir_gap}, where $\alpha=4$, which results in $\misr_{\rm PPP}=1$, while for the triangular
lattice $\misr_{\rm tri}=0.457$. Hence the horizontal shift is $\misr_{\rm PPP}/\misr_{\rm tri}=3.4$ dB.
For Rayleigh fading, we also have the relationship $\ps(\theta)=\calL_{\isr}(\theta)\gtrsim e^{-\theta\,\misr}$
by Jensen's inequality, also shown in the figure. Here '$\gtrsim$' is a lower bound with asymptotic equality. 

In \cite{net:Nigam14tcom}, the authors considered coherent and non-coherent joint transmission for the HIP model
and derived expressions for the SIR distribution. The diversity gain and the asymptotic pre-constants as $\theta\to 0$
are also derived.
In \cite{net:Zhang14twc}, the benefits of BS silencing 
(inter-cell interference coordination) and re-transmissions (intra-cell diversity) in Poisson networks with Rayleigh
fading are studied. For $\theta\to 0$, it is shown that $\ps(\theta) \sim 1-a_k\theta$ when the $k-1$ strongest
interfering BSs are silenced, while $\ps(\theta)\sim 1-b_m\theta^m$ for intra-cell diversity with
$m$ transmissions. For $\theta\to\infty$,
$\ps(\theta)\sim A_k \theta^{-\delta}$ and $\ps(\theta)\sim B_m \theta^{-\delta}$ for BS silencing and retransmissions,
respectively. The constants $a_k$, $b_m$, $A_k$, and $B_m$ are also determined.
Lastly,
\cite[Thm.~2]{net:Miyoshi14aap} gives an expression for the limit $\lim_{\theta\to\infty} \theta^\delta \ps(\theta)$
for the PPP and the Ginibre point process (GPP) with Rayleigh fading. For the GPP,
it consists of a double integral with an infinite product. 

In \cite{net:Blaszczyszyn14arxiv}, the authors consider a Poisson model for the BSs and define a new point process termed {\em signal-to-total-interference-and-noise ratio (STINR) process}.\footnote{What is meant by ``total interference" is actually the total received
power (including the desired signal power).} 
They obtain the moment measures of the new process and use them
to express the probability that the user is covered by $k$ BSs. In our work, we  consider a different map of the original point process based on relative distances, which results in simplified moment measures for the PPP and permits generalizations to other point process models for the base stations.

\subsection{Contributions}
This paper makes the following contributions:
\begin{itemize}
\item We define the {\em relative distance process (RDP)}, which is the relevant point process for cellular networks with
  nearest-BS association, and derive some of its pertinent properties, in particular the probability generating functional (PGFL).
  \item We introduce the {\em generalized MISR}, defined as $\misr_n\triangleq(\E(\isr^n))^{1/n}$, which is applicable to general fading models, and give an explicit expression and tight bounds for the PPP.
\item We provide some evidence why the gain $G_0$ is insensitive to the path loss exponent $\alpha$ and the fading statistics.
\item We show that for 
all stationary point process models and any type of fading, the tail of the
SIR distribution always scales as $\theta^{-\delta}$, \ie, we have
 $\ps(\theta)\sim c\theta^{-\delta}$, $\theta \to \infty$,
  where the constant $c$ captures the effects of the network
 geometry and fading. The asymptotic gain follows as
 \begin{equation}
    G_\infty =\left(\frac{c}{c_{\rm PPP}}\right)^{1/\delta} ,
    \label{g_infty}
\end{equation}
     and we have
 \[ \ps(\theta)\sim \psppp(\theta/G_\infty), \quad\theta\to\infty .\]
 \item We introduce the {\em expected fading-to-interference ratio (EFIR)} and show that the constant $c$
 is related to the EFIR by $c=\efir^\delta$.
  Consequently, $G_\infty$ is 
given by the ratio of the EFIR of the general point process under consideration and the EFIR of the PPP.
 \end{itemize}

\section{System Model}
The base station locations are modeled as a simple stationary  %and motion invariant  spatial 
point process $\Phi \subset \R^2$. Without loss of generality, we assume that the typical user is located at the origin $o$.
The path loss between the typical user and a BS at $x\in \Phi$ is given by $\ell(x)=\|x\|^{-\alpha}$, $\alpha>2$.  
Let $\bar F_h$ denote the ccdf of the iid fading random variables, which are assumed to have mean $1$.

We assume nearest-BS association, wherein  a user is served by the closest BS. Let $x_0$
denote the closest BS to the typical user at the origin and define $R\triangleq \|x_0\|$
and $\Phi^!=\Phi\setminus\{x_0\}$. %More precisely, 
 With the nearest-BS association rule, the downlink SIR \eqref{sir} of the typical user 
 can be expressed as
 \begin{align}
\sir = \frac{hR^{-\alpha}}{\sum_{x\in \Phi^!}h_x\l(x)} .% {I^!},
\label{eq:sir}
\end{align}

{\em Further notation:} $b(o,r)$ denotes the open disk of radius $r$ at $o$,
and $b(o,r)^{\rm c}\triangleq \R^2\setminus b(o,r)$ is its complement. % and let $\Phi^!=\Phi\setminus\{x_0\}$.

\section{The Relative Distance Process}
In this section, we introduce a new point process that is a transformation of the original point process $\Phi$ and helps in the analysis of the  interference-to-signal ratio. 
\subsection{Definition}
\label{sec:rdp_def}
From \eqref{sir}, the MISR is defined as
\begin{equation}
   \misr\triangleq \E\left(\frac{\sum_{x\in\Phi^!}h_x\ell(x)}{\ell(x_0)}\right)= \E_\Phi\left(\frac{\sum_{x\in\Phi^!}\ell(x)}{\ell(x_0)}\right). 
  \label{misr}
\end{equation}
The first expectation is taken over $\Phi$ and $h$, while the second one is only over $\Phi$ since $\E(h)=1$.
Since $\ell(x)$ only depends on $\|x\|$, it is apparent that the MISR is determined by the relative
distances of the interfering and serving BSs. Accordingly, we introduce a new point process on
the unit interval $(0,1)$
that captures only these relative distances.

\begin{definition}[Relative distance process (RDP)]
\label{def:rdp_def}
For a simple stationary point process $\Phi$, let $x_0=\argmin\{x\in\Phi\colon \|x\|\}$. The
{\em relative distance process (RDP)} is defined as
\[ \mathcal{R}\triangleq\{x\in\Phi\setminus\{x_0\}\colon \|x_0\|/\|x\| \}\subset (0,1).\]
\end{definition}
Using the RDP, the $\isr$ can be expressed as
\[ \isr=\sum_{y\in\calR} h_y y^\alpha ,\] %=\sum_{z\in\calP} h_z z ,\]
and, since $\E(h_y)=1$, the MISR is
\begin{align*}
\misr& =\E  \sum_{y \in \calR} y^\alpha
= \int _0^1 r^\alpha \Lambda(\dd r). %\\
\end{align*} 
For the stationary PPP, 
the cdfs of the elements $\nu_k=\|x_0\|/\|x_k\|$ of $\calR$
are $F_{\nu_k}(x)=1-(1-x^2)^k$,
$x\in [0,1]$, as given in  \cite{net:Haenggi14wcl}. Summing the densities $\dd F_{\nu_k}(x)/\dd x$ over $k\in\N$ yields
the intensity measure $\Lambda(\dd r)=2r^{-3}\dd r$. 
It follows that the mean measure $\Lambda([r,1))\triangleq\E\calR([r,1))=r^{-2}-1$, $0<r<1$. 
The fact that the mean measure diverges near $0$ is consistent with the fact that $\calR$ is not
locally finite on intervals $[0,\epsilon)$.

Generally, since the MISR only depends on the relative distances, the gain $G_0$ does not depend on the
base station density.

\iffalse % not interesting at this point.
\begin{definition}[Relative path loss process (RPLP)]
For an RDP $\calR$, the {\em relative path loss process (RPLP)} is defined as
\[ \calP\triangleq \{r\in\calR\colon r^\alpha \}\subset (0,1).\]
\end{definition}
For a PPP, the corresponding RPLP has intensity $\Lambda(\dd r)=2 r^{\alpha-3}$, $r\in(0,1)$.
So $\alpha=3$ is the neutral case, where the RPLP is stationary on $(0,1)$.
\fi

\subsection{RDP of the PPP}
The success probability for Rayleigh fading is given by the Laplace transform of the $\isr$:
\begin{equation}
  \ps(\theta)=\E e^{-\theta\,\isr}=\E\prod_{y\in\calR} e^{-\theta h_y y^\alpha}=\E\prod_{y\in\calR} \frac{1}{1+\theta y^\alpha}
  \label{ps_rdp_rayl}
\end{equation}
This RDP-based formulation has the advantage that it circumvents the usual two-step procedure, where first the
conditional success probability given the distance to the serving base station $R$ is calculated and then an
expectation with respect to $R$ is taken.

\iffalse %%%%%%%%%%%%%%
\rk{--- Updated----}

Let $\Delta(y)=\frac{  y^{\alpha}}{1+\theta y^{\alpha}}$. The success probability equals
\[\ps(\theta)= \E\prod_{y\in \calR} 1- \theta \Delta(y), \]
expanding which we obtain the following lower bound
\begin{align*}
\ps(\theta)&\geq 1- \theta \E \sum_{y\in \calR}  \Delta(y)\\
& = 1- 2\theta\int_0^1\Delta(r)r^{-3}\dd r \\
&=1-\frac{\delta \theta  \, _2F_1\left(1,1-\delta;2-\delta;-\theta
   \right)}{1-\delta} \\
   &= 2-\frac1{\ps(\theta)}.
\end{align*}
The last step follows from \eqref{ps_ppp} and the identity
\[ \frac{\delta \theta  \, _2F_1\left(1,1-\delta;2-\delta;-\theta
   \right)}{1-\delta} \equiv \,_2F_1(1,-\delta; 1-\delta; -\theta)-1 .\]
\fi
Since \eqref{ps_rdp_rayl} has the form of a PGFL, we first 
calculate the PGFL of the RDP generated by a PPP.
\begin{lemma}
\label{lem:pgfl_rdp}
When $\Phi$ is a PPP, the probability generating  functional of the RDP is given by
\begin{align}
G_{\calR}[f]\triangleq \E\prod_{x\in \mathcal{R}} f(x) = &\frac{1}{1+2\int_0^1 (1-f(x))x^{-3}\dd x},
\label{rdp_ppp_pgfl}
\end{align}
for functions $f\colon [0,1]\mapsto[0,1]$ such that the integral in the denominator of \eqref{rdp_ppp_pgfl} is finite. 
\end{lemma}
\begin{IEEEproof}
The PGFL $G_{\calR}[f]$ can be calculated as
\begin{align*}
G_{\calR}[f] &= \E\left[\prod_{x\in \mathcal{R}} f(x)\right]\\
&= \E\left[\prod_{x\in \Phi\setminus\{x_0\}} f\left (\frac{\|x_0\|}{\|x\|}\right)\right]\\
&\stackrel{(a)}{=}\E_R \exp\left(-2 \pi \lambda\int_R^\infty a \left(1-f\left (\frac{R}{a}\right)\right)\dd a\right)\\
&\stackrel{(b)}{=}\lambda\int\limits_0^\infty 2\pi r \exp\left(-2 \pi \lambda\int\limits_r^\infty a \left(1-f\left (\frac{r}{a}\right)\right)\dd a\right)e^{-\lambda \pi r^2}\dd r\\
&=\lambda\int\limits_0^\infty 2\pi r \exp\left(-2 \pi r^2 \lambda\int\limits_1^\infty y \left(1-f\left (\frac{1}{y}\right)\right)\dd y\right)e^{-\lambda \pi r^2}\dd r\\
&=\frac{1}{1+2\int_1^\infty y \left(1-f(1/y)\right)\dd y},
\end{align*}
where $(a)$ is obtained from the PGFL of the PPP, which, for
a general PPP $\Psi$ with intensity measure $\Lambda$, is given by
\begin{equation}
 G_\Psi[f]\triangleq \E\prod_{x\in\Psi} f(x)=\exp\left(-\int_{\R^2} [1-f(x)]\Lambda(\dd x)\right) .
 \label{ppp_pgfl}
 \end{equation}
Writing the PGFL in polar form
and conditioning on the distance to the nearest neighbor $R=\|x_0\|$ yields $(a)$.
In $(b)$, we de-condition on $R$ using the
the nearest-neighbor distribution of the PPP.
Using the substitution $y^{-1}\to x$, we obtain the final result.
\end{IEEEproof}
When $f(x)=1/(1+\theta x^{\alpha})$ (see \eqref{ps_rdp_rayl}),
we retrieve the result in \eqref{ps_ppp} for Poisson cellular networks with Rayleigh fading.

It may be suspected that the RDP of a PPP is itself a (non-stationary) PPP on $[0,1]$. It is easily
seen that this is not the case. Let $\Psi$ be a PPP on $[0,1$] with the same intensity function as $\calR$, \ie,
$\Lambda(\dd r)=2r^{-3}\dd r$. If $\calR$ was a PPP,
the success probability for Rayleigh fading would follow from
the PGFL of $\Psi$ (specializing \eqref{ppp_pgfl} to a PPP on $[0,1]$) as
\[ G_\Psi[f]=\exp\left(-\int_0^1 [1-f(r)] 2r^{-3}\dd r\right) ,\]
and, in turn, the success probability 
would be given by
\begin{equation}
   \tps(\theta)=\exp\left(-\int_0^1 \frac{\theta r^\alpha}{1+\theta r^\alpha} 2r^{-3}\dd r\right)
   \label{rdp_ppp}
 \end{equation}
 instead of \eqref{ps_ppp}. 

 \arxiv
 However, assuming $\calR$ to be Poisson yields an approximation of the
success probability, with asymptotic equality as $\theta\to 0$. The ``Poisson approximation" \eqref{rdp_ppp}
is related to the actual value \eqref{ps_ppp} as
\[ \tps(\theta)=\exp\left(1-\frac1{\ps(\theta)}\right) ,\]
This holds due to the identity
\[ \frac{\delta \theta  \, _2F_1\left(1,1-\delta;2-\delta;-\theta
   \right)}{1-\delta} \equiv \,_2F_1(1,-\delta; 1-\delta; -\theta)-1 .\]
Rewriting and expanding, we have 
\[ \tps(\theta)=\frac{1}{\exp(1/\ps(\theta)-1)} =\frac{1}{\frac1{\ps(\theta)}+\frac12\left(\frac1{\ps(\theta)}-1\right)^2+\ldots}.\]
Hence, only considering the dominant first term in the denominator as $\theta\to 0$,
we obtain $\tps(\theta)\sim \ps(\theta)$, $\theta\to 0$.
\fi

The fact that $\tps(\theta)<\ps(\theta)$ for $\theta>0$ is an indication that the higher moment densities of the RDP are larger
than those of the PPP. This is indeed the case, as the following calculation of the moment densities shows.

\begin{lemma}
\label{lem:moment_densities_rdp}
When $\Phi$ is a PPP, the moment densities of the RDP are given by
\begin{align}
\rho^{(n)}(t_1,t_2, \hdots, t_n) = n!\,2^n\prod_{i=1}^n t_i^{-3}.
\label{rdp_rho}
\end{align}
 \end{lemma}
\begin{IEEEproof}
First we obtain the factorial moment measures. % of the RDP. 
We use the simplified notation\footnote{Here the
intervals are chosen as $(t,1)$ for $t>0$ since the RDP is not locally finite on $[0,\epsilon]$.} 
\begin{equation*}
   \alpha^{(n)}(t_1,t_2, \hdots, t_n)\equiv  \alpha^{(n)}((t_1,1)\times(t_2,1)\times \hdots\times (t_n,1)),\quad 0<t_i\leq 1 .
\end{equation*}
The factorial moment measures are defined as
\begin{equation*}
 \alpha^{(n)}(t_1,t_2, \hdots, t_n) %&\triangleq  \alpha^{(n)}((t_1,1),(t_2,1), \hdots, (t_n,1)),\quad 0<t_i\leq 1, \\
  = \E\sum_{x_1,x_2,\hdots,x_n\in\calR}^{\neq} \mathbf{1}_{(t_1,1)}(x_1)\mathbf{1}_{(t_2,1)}(x_2)\cdots \mathbf{1}_{(t_n,1)}(x_n), 
\end{equation*}
where $\sum^{\neq}$ indicates that the sum is taken over $n$-tuples of {\em distinct} points.
The moment measures are related to the PGFL as \cite[p.~116]{net:Stoyan95}
\begin{equation}
 \alpha^{(n)}(t_1,t_2, \hdots, t_n)\equiv  (-1)^n\frac{\partial }{\partial s_1}\hdots \frac{\partial }{\partial s_n}G_{\calR}[1-s_1\mathbf{1}_{(t_1,1)}-s_2\mathbf{1}_{(t_2,1)}-\hdots - \mathbf{1}_{(t_n,1)}]
 \label{mom_measures_pgfl}
 \end{equation}
evaluated at $s_1=s_2=\hdots=s_n=0$. Using Lemma \ref{lem:pgfl_rdp} we obtain 
\begin{equation*}
 G_{\calR}[1-s_1\mathbf{1}_{(t_1,1)}-s_2\mathbf{1}_{(t_2,1)}-\hdots - s_n\mathbf{1}_{(t_n,1)}]= \\
 \frac{1}{1+\sum_{i=1}^n s_i(t_i^{-2}-1)}.
\end{equation*}
Differentiating with respect to $s_i$ and setting $s_1=s_2=\cdots=s_n=0$, we have
\begin{align}
\alpha^{(n)}(t_1,t_2, \hdots, t_n) = n!\prod_{i=1}^n\left(\frac{1}{t_i^2}-1\right).
\end{align}
The moment densities follow from differentiation, noting that $t_i$ denotes the start of the interval, which causes
a sign change since increasing $t_i$ decreases the measure.
\end{IEEEproof}
So the product densities are a factor $n!$ larger than they would be if $\calR$ was a PPP. This implies,
interestingly, that the pair correlation function \cite[Def.~6.6]{net:Haenggi12book} of the RDP
of the PPP is $g(x,y)=2$, $\forall x,y\in(0,1)$. 

\arxiv
The moment densities of the RDP provide an alternative way 
to obtain the success probability for the PPP:
\begin{align*}
  \ps(\theta)&= \E\prod_{y\in\calR} \frac{1}{1+\theta y^\alpha} = \E  \prod_{y\in\calR} \left(1-\frac{1}{1+\theta^{-1} y^{-\alpha}}\right)\\
&=1-\sum_{y\in\calR} \nu(\theta,y)+\frac{1}{2!}\E\sum^{\neq}_{y_1,y_2\in\calR}\nu(\theta,y_1)\nu(\theta,y_2)+\hdots +\frac{(-1)^n}{n!}\E\sum^{\neq}_{y_1,\hdots,y_n\in\calR}\prod_{i=1}^n\nu(\theta,y_i)+\hdots,
\end{align*}
where $\nu(\theta,y)=\frac{1}{1+\theta^{-1} y^{-\alpha}}$.
From the definition of the moment densities, we have
\begin{align}
 \ps(\theta)&= \sum_{n=0}^\infty \frac{(-1)^n}{n!}\E\sum^{\neq}_{y_1,\hdots,y_n\in\calR}\prod_{i=1}^n\nu(\theta,y_i)\nonumber\\
 &= \sum_{n=0}^\infty \frac{(-1)^n}{n!}\int_{[0,1]^n} \left(\prod_{i=1}^n\nu(\theta,t_i) \right)\rho^{(n)}(t_1,t_2, \hdots, t_n)\dd t_1\hdots \dd t_n
 \label{eq:pc_moment_measure}
\end{align}
Using Lemma \ref{lem:moment_densities_rdp}, we have
\begin{align*}
 \ps(\theta)&=  \sum_{n=0}^\infty  {2^n(-1)^n} \left(\int_{[0,1]}\nu(\theta,t)  t^{-3} \dd t \right)^n,\\
 &=\sum_{n=0}^\infty  { \theta^n(-1)^n}\left( \frac{  \delta \, _2F_1\left(1,1-\delta;2-\delta;-\theta
   \right)}{1 -\delta}\right)^n\\
   &=\sum_{n=0}^\infty  { \theta^n(-1)^n}\left( \,_2F_1(1,-\delta; 1-\delta; -\theta)-1\right)^n\\
   &= \frac{1}{\,_2F_1(1,-\delta; 1-\delta; -\theta)},
\end{align*}
 which equals the success probability given in  \eqref{ps_ppp}.
\fi

\subsection{RDP of a stationary  point process}
We now characterize the PGFL of the RDP generated by a stationary point process. Let $f(R, \Phi^!)$ be a positive function of the distance $R=\|x_0\|$ and the point process $\Phi^!=\Phi\setminus \{x_0\} $. % real-valued
The average $\E[f(R, \Phi^! )]$ can in principle be evaluated using the joint distribution of $R$ and $\Phi$,
which is, however, known only for a few point processes.
Thus we introduce an alternative representation of $f(R,\Phi^!)$ that is easier to work with. 

The indicator variable $ \one(\Phi(b(o,\|x\|))=0)$, $x\in \Phi$, equals one only when  $x=x_0$.
Hence it follows that
\begin{align}
f(R, \Phi^!) \equiv \sum_{x\in \Phi} f(\|x\|,\Phi \setminus \{x\} ) \one(\Phi(b(o,\|x\|))=0).
\label{eq:rep}
\end{align}
This representation of $f(R, \Phi^!)$ permits the computation of the expectation of $f(R,\Phi^!)$  using the Campbell-Mecke theorem
\cite[Thm.~8.2]{net:Haenggi12book}.
We use the above idea in the next lemma to obtain the PGFL of a general RDP.

 \begin{lemma}
\label{lem:rdp_general}
The PGFL of the RDP generated by a stationary point process $\Phi$  is given by
\begin{align}
G_{\cal R}[f]=\lambda\int_{\R^2 }\calG^{!}_o\left[ f\left (\frac{\|x\|}{\|\cdot +x\|}\right)\i(\cdot +x \in b(o,\|x\|)^{\rm c} )\right] \dd x,
\label{rdp_gen_pgfl}
\end{align}
where $\calG^{!}_o$ is the PGFL of the point process $\Phi$ with respect to the reduced Palm measure. 
\end{lemma}
\begin{IEEEproof}
Using the representation in \eqref{eq:rep}, we obtain
\begin{align*}
G_{\calR}[f] &=\E\sum_{x\in \Phi} \prod_{y\in \Phi \setminus \{x\} } f\left (\frac{\|x\|}{\|y\|}\right)\i(\Phi(b(o,\|x\|))=0) \\
 &\stackrel{{\rm (a)}}= \lambda\int_{\R^2} \E_o^! \prod_{y\in \Phi } f\left (\frac{\|x\|}{\|y+x\|}\right)\i(\Phi(b(-x,\|x\|))=0) \dd x\\
&=\lambda\int_{\R^2} \E_o^! \prod_{y\in \Phi } f\left (\frac{\|x\|}{\|y+x\|}\right)\i(y\in b(-x,\|x\|)^{\rm c} ) \dd x\\
&=\lambda\int_{\R^2} \mathcal{G}_o^!\left[ f\left (\frac{\|x\|}{\|\cdot +x\|}\right)\i(\cdot +x \in b(o,\|x\|)^{\rm c} )\right] \dd x,
\end{align*}
where $(a)$ follows from the Campbell-Mecke theorem.
\end{IEEEproof}
If $\Phi$ is  also rotationally invariant (\ie, motion-invariant), the reduced Palm measure is also rotationally invariant  and hence
\begin{align}
G_{\cal R}[f] = \lambda2\pi \int\limits_{0}^\infty r \mathcal{G}_o^!\left[ f\left (\frac{r}{\|\cdot +r\|}\right)\i(\cdot +r \in b(o,r)^{\rm c} )\right] \dd r,
\label{eq:rot_inv}
\end{align}
where $r$ in any vector addition should be interpreted as $(r,0)$. 

{\em Remark.} Taking $\Phi$ to be a PPP, \eqref{eq:rot_inv} reduces to \eqref{rdp_ppp_pgfl} since
\begin{align*}
G_{\calR}[f]&=2\pi\lambda\int_0^\infty r\exp\Bigg(-\lambda\int_{\R^2} \bigg[1-f\left(\frac{r}{\|x+(r,0)\|}\right)
 \i(x+(r,0)\in b(0,r)^{\rm c})\bigg]\dd x \Bigg)\dd r\\
&=2\pi\lambda\int\limits_0^\infty r\exp\bigg(-\lambda\pi r^2-\lambda\int\limits_{b(o,r)^c}\left[1-f\left(\frac{r}{\|x\|}\right)\right]\dd x\bigg)\dd r\\
&=2\pi\lambda\int_0^\infty r\exp\left(-\lambda\pi r^2-2\pi\lambda\int_r^\infty y[1-f(r/y)]\dd y\right)\dd r.
\end{align*}
The last expression equals the second-to-last line in the proof of Lemma \ref{lem:pgfl_rdp}.

Similar to the case of PPP, we now obtain the moment measures of the RDP of a general stationary point process.
 \begin{lemma}
\label{lem:rho_gen}
The factorial moment measures 
of the RDP $\calR$ generated by a stationary point process $\Phi$ are
\begin{equation}
\alpha^{(n)}(t_1,t_2, \hdots, t_n) =   \int\limits_{\R^{n+1}} \prod_{i=1}^n f(y_0,y_i,t_i) \E^{!}_{y_0,y_1,\hdots,y_n}[g(\Phi,y_0)] 
 \rho^{(n+1)}_\Phi(y_0,y_1,\hdots,y_n) \dd y_0\hdots \dd y_n,
\label{fact_rdp}
\end{equation}
where 
$g(\Psi,x) =  \prod_{y\in \Psi  }  \i(\|y\|\geq \|x\|)$ and  $f(x,y,t) = \mathbf{1}_{(t,1)}\left (\frac{\|x\|}{\|y\|}\right)$. 
The product densities are
\begin{align}
\rho^{(n)}_{\calR}(t_1,t_2, \hdots, t_n) =   n!\,2^n\left( \prod_{i=1}^n t_i^{-3}\right) \beta_n(t_1,\hdots, t_n),
\label{densities_rdp}
\end{align}
where 
\begin{align}
\beta_n(t_1,\hdots, t_n)\triangleq&\frac{1}{n!2^n}\int\limits_{\R^2}  \|y_0\|^{2n}  \int\limits_{[0, 2\pi]^n} \E^{!}_{y_0,\left(\frac{\|y_0\|}{t_1},\varphi_1\right) ,\hdots,\left(\frac{\|y_0\|}{t_n},\varphi_n\right)}[g(\Phi,y_0)]\nonumber\cdot \\
 &\;\rho^{(n+1)}_\Phi\left(y_0,\Big(\frac{\|y_0\|}{t_1},\varphi_1\Big),\ldots, \Big(\frac{\|y_0\|}{t_n},\varphi_n\Big)\right) \dd \varphi_1\hdots \dd \varphi_n  \dd y_0.
 \label{beta_n}
 \end{align}
  \end{lemma}
\begin{IEEEproof}
As before, we use the relationship \eqref{mom_measures_pgfl}.
While the result can be obtained from the PGFL in Lemma \ref{lem:pgfl_rdp}, it is easier to begin with the definition of the PGFL. 
We have 
\begin{align*}
G_{\calR}[f] &=\E\sum_{x\in \Phi} \prod_{y\in \Phi \setminus \{x\} } f\left (\frac{\|x\|}{\|y\|}\right)\i(\|y\|\geq \|x\|)\\
&=\E\sum_{x\in \Phi} g(\Phi\setminus \{x\},x)  \prod_{y\in \Phi \setminus \{x\} } f\left (\frac{\|x\|}{\|y\|}\right).
\end{align*}
We are interested in the derivative of the PGFL with  the function $f(z) = 1- \sum_{i=1}^n s_i \i_{(t_i,1)}(z)$. So we have
\begin{equation*}
\alpha^{(n)}(t_1,t_2, \hdots, t_n)
= (-1)^n\frac{\partial }{\partial s_1}\hdots \frac{\partial }{\partial s_n}\E\sum_{x\in \Phi} g(\Phi\setminus \{x\},x) 
 \prod_{y\in \Phi \setminus \{x\} } \left(1- \sum_{i=1}^n s_i \i_{(t_i,1)}\left(\frac{\|x\|}{\|y\|}\right) \right)
\end{equation*}
evaluated at $s_1=s_2=\hdots=s_n=0$.
 Expanding the inner product over the summation we obtain an infinite polynomial in the powers of $s_1,\hdots, s_n$ and their products.  We observe that the only term that contributes to the derivative in a non-zero manner  is the $s_1 s_2\cdots s_n$ term.
 Its coefficient equals
 \begin{equation*}
  T=(-1)^n \E \bigg(\sum_{x\in\Phi}g(\Phi\setminus \{x\},x)
    \sum_{ y_1, \hdots y_n \in \Phi \setminus \{x\}}^{\neq} f(x,y_1,t_1)\hdots f(x,y_n,t_n)\bigg). 
 \end{equation*}
 Combining the summations,
 \begin{equation*}
  T=(-1)^n \E  \bigg( \sum_{ y_0,y_1, \hdots y_n \in \Phi}^{\neq} f(y_0,y_1,t_1)\cdot \hdots \cdot 
    f(y_0,y_n,t_n)g(\Phi\setminus \{y_0\},y_0)\bigg).
 \end{equation*}
Since $f(y_0,y_i,t_i) \neq 1$ implies $\|y_i \|\geq \|y_0\|$, %the above expression equals
\begin{equation*}
T=  (-1)^n \E  \bigg( \sum_{ y_0,y_1, \hdots y_n \in \Phi}^{\neq} f(y_0,y_1,t_1)\cdot \hdots\cdot
   f(y_0,y_n,t_n)g(\Phi\setminus \{y_0,y_1,\hdots, y_n\},y_0)\bigg),
\end{equation*}
and, using \cite[Thm.~1]{net:Hanish1982},
\begin{equation}
 T=(-1)^n \int_{\R^{n+1}} \prod_{i=1}^n f(y_0,y_i,t_i) \E^{!}_{y_0,y_1,\hdots ,y_n}[g(\Phi,y_0)] \cdot \\
  \rho^{(n+1)}_\Phi(y_0,y_1,\hdots,y_n) \dd y_0 \hdots \dd y_n,
\end{equation}
and the result \eqref{fact_rdp} follows.
For the product densities, we convert the variables $x_i$ into polar
coordinates $(r_i,\varphi_i)$,  which yields %we have  $\alpha^n(t_1,t_2,\hdots,t_n)=$,
\begin{align*}
\alpha^n(t_1,t_2,\hdots,t_n) &= \int\limits_{\R^2} \int\limits_{\|y_0\|}^{\|y_0\|/t_1}\!\! r_1 \cdots \!\! \int\limits_{\|y_0\|}^{\|y_0\|/t_n}r_n \!\! \int\limits_{[0, 2\pi]^n} \!\!
\E^{!}_{y_0,(r_1,\varphi_1) ,\hdots,(r_n,\varphi_n)}[g(\Phi,y_0)]\nonumber \\
 &\qquad\cdot \rho^{(n+1)}(y_0,(r_1,\varphi_1),.., (r_n,\varphi_n)) \dd \varphi_1 \hdots \dd \varphi_n \dd r_1\hdots \dd r_n \dd y_0.
\end{align*}
Then differentiating using the Leibniz rule with respect to $t_1, \hdots, t_n$, we obtain 
\begin{multline*}
\rho^{(n)}_{\calR}(t_1,t_2, \hdots, t_n)
= \left( \prod_{i=1}^n t_i^{-3}\right)\int\limits_{\R^2}  \|y_0\|^{2n} 
 \int\limits_{[0, 2\pi]^n} \E^{!}_{y_0,\left(\frac{\|y_0\|}{t_1},\varphi_1\right) ,\hdots,\left(\frac{\|y_0\|}{t_n},\varphi_n\right)}[g(\Phi,y_0)] \cdot \\
 \rho^{(n+1)}_\Phi\left(y_0,\left(\frac{\|y_0\|}{t_1},\varphi_1\right),\ldots, \left(\frac{\|y_0\|}{t_n},\varphi_n\right)\right) \dd \varphi_1 \hdots \dd \varphi_n \dd y_0,
\end{multline*}
which equals \eqref{densities_rdp}.
\end{IEEEproof}
\arxiv
As in \eqref{eq:pc_moment_measure}, the moment densities of the RDP generated by a stationary point process can be used to compute its  corresponding success probability.  
\fi

\section{The Mean Interference-to-Signal Ratio (MISR) and the Gain at $0$}
In this section, we introduce and analyze the MISR, including its generalized version, and apply it to
derive a simple asymptotic expression of the SIR distribution near $0$ using the gain $G_0$. We
also give some insight why $G_0$ barely depends on the path loss exponent $\alpha$ and the fading statistics.
\subsection{The MISR for general point processes}
The first result gives an expression for the MISR for a general point process. 
 \begin{theorem}
\label{thm:misr_pp}
The MISR of a motion-invariant point process $\Phi$ is given by
\[\misr =  2\int_0^1 t^{\alpha -3}\beta_1(t)\dd t,\]
where $\beta_1(t)$ is given in \eqref{beta_n} in Lemma \ref{lem:rho_gen}.
 \end{theorem}
\begin{IEEEproof}
Using the RDP, the MISR can be expressed as
\begin{align*}
\misr& =\E  \sum_{y \in \calR} y^\alpha\\
& = \int _0^1 t^\alpha \rho^{(1)}_{\calR}(t)\dd t.\\
&\stackrel{(a)}{=} 2 \int _0^1 t^\alpha t^{-3} \beta_1(t)\dd t,
 \end{align*} 
 where $(a)$ follows from Lemma \ref{lem:rho_gen}.
  \end{IEEEproof}
When $\Phi$ is a PPP, from Slivnyak's theorem and the fact that $\rho^{(2)} = \lambda^2$, we have
$\beta_1(t) = 1$ and hence $\misr = 2/(\alpha -2)$.

\subsection{The Generalized MISR}
\begin{definition}[Generalized MISR]
The {\em generalized MISR} with parameter $n$ is defined as
\[ \misr_n\triangleq (\E(\isr^n))^{1/n} .\]
\end{definition}
If there is a danger of confusion, we call $\misr\equiv\misr_1$ the {\em standard MISR}.

The generalized MISR can be obtained by taking the corresponding derivative of the 
Laplace transform $\E(e^{-s\,\isr})$ at $s=0$. In case of the PPP with Rayleigh fading,
the Laplace transform is known and equals the success probability \eqref{ps_ppp},
thus
\begin{equation}
 \misr_n^n=\E(\isr^n) = (-1)^n \frac{\dd}{\dd \theta^n} \ps(\theta)\big|_{\theta=0}.
 \label{gen_misr_ppp_rayl}
 \end{equation}
For general fading, the Laplace transform is not known, but we can still calculate the
derivative at $s=0$, as the following result
for the PPP with general fading shows.

\begin{theorem}[Generalized MISR and lower bound for PPP]
\label{thm:gen_misr_ppp}
For a Poisson cellular network with arbitrary fading,
\begin{equation}
 \E(\isr^n)=\sum_{k=1}^n k! B_{n,k}\left(\frac{\delta}{1-\delta},\hdots ,\frac{\delta\E(h^{n-k+1})}{n-k+1-\delta}\right),
 \label{gen_misr_ppp}
\end{equation}
where $B_{n,k}$ are the (incomplete) Bell polynomials.
For $n>1$, the generalized MISR is lower bounded as
\begin{equation}
  \misr_n \geq \left[\left(\frac{\delta}{1-\delta}\right)^n n! + \frac{\delta\E(h^n)}{n-\delta}\right]^{1/n} .
  \label{misr_bound_ppp}
\end{equation}
For $n=2$, equality holds, and for $\delta\to 0$ and $\delta\to 1$, the lower bound is asymptotically tight.
\end{theorem}
\begin{IEEEproof}
  We begin with the the Laplace transform of the $\isr$, given by
\begin{align*}
\E(e^{-s\,\isr}) &= \E \prod_{y\in \calR} \L_h(sy^\alpha)\\
&\stackrel{(a)}{=} \frac{1}{1+2\int_1^\infty y \left(1-\L_h\left (\frac{s}{y^\alpha}\right)\right)\dd y}.
\end{align*}
where $(a)$ follows from Lemma \ref{lem:pgfl_rdp}. 
Let $f(s)= 1/(1+s)$ and $g(s)=2\int_1^\infty y \left(1-\L_h\left (\frac{s}{y^\alpha}\right)\right)\dd y$. Then $\E(e^{-s\,\isr})  = f(g(s))$.  We are interested in the $m$-derivative of $\E(e^{-s\,\isr})$ with respect to $s$ at $s=0$, which can be computed using
Fa\`a di Bruno's formula \cite{net:Johnson02amm} as
\begin{equation*}
  \frac{\dd }{\dd s}f(g(s))|_{s=0} = \\ \sum_{k=1}^n  f^{(k)}(g(0))B_{n,k}(g'(0),g''(0),...,g^{(n-k+1)}(0)),
\end{equation*}
where $B_{n,k}$ are the (incomplete) Bell polynomials. 
We have $g(0)=0$, 
\[f^{(k)}(s)= (-1)^{k}\frac{k!}{(1+s)^{k+1}},\]
and
\[g^{(k)}(s) = -2\int_1^\infty y^{1-k\alpha} \L_h^{(k)}(s y^{-\alpha})\dd y,\]
which, when evaluated at $s=0$, equals
\[g^{(k)}(0) = \frac{2(-1)^{k+1}\E(h^k)}{k\alpha-2}.\]
Combining everything, we have
\begin{align}
\E(\isr^n)&=(-1)^n \frac{\dd^n }{\dd s^n } \E(e^{-s\,\isr}) \big|_{s=0} \nonumber \\
 &=  (-1)^n\sum_{k=1}^n(-1)^k k!  B_{n,k}\left(\frac{2}{\alpha-2},\hdots ,\frac{2(-1)^{n-k}\E(h^{n-k+1})}{(n-k+1)\alpha-2}\right).
\label{eq:bell_1}
\end{align}
From the definition of Bell polynomials it follows that all the terms are positive, hence the result
\eqref{gen_misr_ppp} follows from $\delta=2/\alpha$.  The lower bound is obtained by only considering the
terms $k=1$ and $k=m$ in the sum \eqref{eq:bell_1}.
The bound becomes tight as $\delta\to 0$ {\em and} as $\delta\to 1$ since the term $k=1$ dominates the sum \eqref{eq:bell_1}
as $\delta\to 1$
since it is the only term with a denominator $(1-\delta)^n$, 
while the term $k=n$ dominates as $\delta \to 0$ since it is the only one with a numerator $\Theta(\delta)$.
\end{IEEEproof}
Hence we have two simpler asymptotically tight bounds for the generalized MISR:
\begin{align}
  \misr_n &\gtrsim \left(\frac{\delta}{n} \E(h^n)\right)^{1/n},\quad \delta\to 0 \label{asym_delta0} \\
 \misr_n & \gtrsim \frac{\delta(n!)^{1/n}}{1-\delta}=\misr_1 (n!)^{1/n},\quad \delta\to 1. \label{asym_delta_inf}
\end{align}
For Rayleigh fading, \eqref{asym_delta0} yields
$\misr_n \sim (\delta\Gamma(n))^{1/n}$, $\delta\to 0$.

\begin{figure}
\centerline{\includegraphics[width=\figwidth]{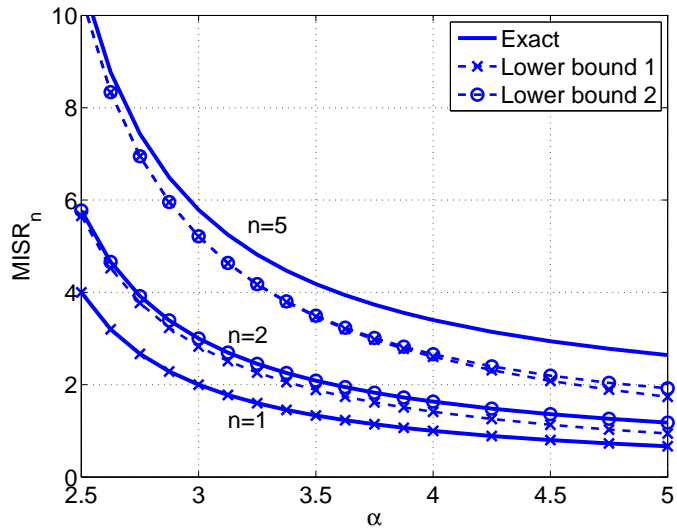}} %{MISR_versus_alpha_bound}}
\caption{ $\misr_n$ for $n\in\{1,2,5\}$ for the PPP with Rayleigh fading as a function of the path loss exponent $\alpha$.
``Lower bound 1" is the 
simple bound in \eqref{asym_delta_inf}, which holds irrespective of the fading and is asymptotically tight as
$\alpha\downarrow 2$, and ``Lower bound 2" is the bound
in \eqref{misr_bound_ppp}, which is valid for $n\geq 2$ and is exact for $n=2$. For $\alpha\leq 4$, the
two bounds are essentially identical.}
\label{fig:MISR_alpha}
\end{figure}

\figref{fig:MISR_alpha} shows $\misr_n$ for Rayleigh fading as a function of the path loss exponent.
As can be observed, the term $\misr_1 (n!)^{1/n}$ is dominant for $\alpha\leq 4$ even if the fading is severe (Rayleigh
fading). For less severe fading, the term with $\E(h^n)$ is less relevant; it only becomes dominant for
unrealistically high path loss exponents ($\delta\ll 1$).

The second moment of the $\isr$ follows from \eqref{gen_misr_ppp} as
\[ \E(\isr^2)=2\,\misr_1^2+\frac{\delta\E(h^2)}{2-\delta}, \] % \frac{8}{(\alpha-2)^2}+\frac{\E(h^2)}{\alpha-1}
and the third moment is
\[ \E(\isr^3)=6\,\misr_1^3+ \frac{6\delta^2\E(h^2)}{(1-\delta)(2-\delta)}+\frac{\delta\E(h^3)}{3-\delta} .\] %=\frac{2\E(h^3)}{3\alpha-2}+\frac{24\E(h^2)}{(\alpha-2)(2\alpha-2)}+\frac{48}{(\alpha-2)^3} .\]

\arxiv
{\em Remarks.} 
\begin{itemize}
\item Setting $\E(h^k)=1$ for all $k$ retrieves the result in \cite[Prop.~3]{net:Zhang14twc} on the
pre-constant for $m$ transmissions in a Poisson networks over Rayleigh fading.

\iffalse
Again the last term is $n! \,\misr_1^n$, as asserted above.
\mh{Can we improve upon the bound by, say, considering another term?}
\rk{We  can add the first term corresponding to $k=1$} 
Previously we had the following bound
\[\misr_n > \misr_1(n!)^{1/n} = \left(\left(\frac{2\E[h]}{\alpha -2}\right)^n n!\right)^{1/n}.\]
Adding the first term (corresponding to $k=1$) in \eqref{eq:bell_1}, we have
\[\misr_n >  \left(\left(\frac{2\E[h]}{\alpha -2}\right)^n n! + \frac{2 \E[h^n]}{n\alpha-2}\right)^{1/n}.\]
This is an asymptotically tight bound as $\alpha\to \infty$. 
With Rayleigh fading the bound equals
\[\misr_n >  \left(\left(\frac{2\E[h]}{\alpha -2}\right)^n + \frac{2}{n\alpha-2}\right)^{1/n} (n!)^{1/n}.\]
This result also shows \rk{have to prove} that
\begin{align}
\misr_n \sim \left(\frac{2\E[h^n]}{n}\right)^{1/n} \alpha^{-1/n}, \quad \alpha \to \infty.
\label{eq:asymp_misr}
\end{align}
\mh{Actually the improved bound is tight on both ends.}
Writing the improved bound in terms of $\delta=2/\alpha$ (and setting $\E(h)=1$), we have
\[ \E(\isr^n) \geq \left(\frac{\delta}{1-\delta}\right)^n n! + \frac{\delta\E(h^n)}{n-\delta} .\]
This bound becomes tight as $\delta\to 0$ {\em and} as $\delta\to\infty$:
\begin{align}
  \misr_n &\sim \left(\frac{\delta}{n} \E(h^n)\right)^{1/n},\quad \delta\to 0 \label{asym_delta0} \\
 & \sim \frac{(n!)^{1/n}}{1-\delta},\quad \delta\to 1.
\end{align}
\fi

\item
An alternative way to derive the lower bound is as follows.
Letting $u_y\triangleq h_y y^{\alpha}$ for $y\in\calR$, we expand $\isr^n$ as
\begin{align*}
   \isr^n&=\left( \sum_{y\in\calR} u_y\right)^n \\
   &=\sum_{y\in\calR} u_y^n +\binom n1\sum_{y\in\calR} \left(u_y^{n-1} \sum_{x\in\calR\setminus\{y\}} u_x\right) + 
       \binom n2\sum_{y\in\calR} \left(u_y^{n-2} \sum_{x,z\in\calR\setminus\{y\}} u_x u_z\right)+\\
       &\;\;\binom n3\sum_{y\in\calR} \left(u_y^{k-3} \sum_{v,x,z\in\calR\setminus\{y\}} u_v u_x u_z\right)+
    \underbrace{\ldots}_{n-4 \text{ terms}} +\binom nn\sum_{x_1,\ldots,x_n\in\calR}^{\neq} u_{x_1} \cdots u_{x_n} ,
\end{align*}
where the expression contains $k$ sums. Ignoring all but the first and last terms of the expansion, we obtain
\begin{align*}
  \E(\isr^n) &\geq \E(h^n)\frac{\delta}{m-\delta} +\int_{[0,1]^n} (x_1\cdots x_n)^{\alpha} \rho_\Phi^{(n)}(x_1,\ldots,x_n) \dd x_1\cdots\dd x_n \\
   &=\E(h^n)\frac{\delta}{n-\delta} + n!\, \misr_1^n ,
\end{align*}
which equals the result in \eqref{misr_bound_ppp}.
\end{itemize}
\fi

For Nakagami-$m$ fading, $\misr_n$ is decreasing with increasing $m$ since the moments $\E(h^n)$ are decreasing
with $m$. As the lower bound $\misr_1 (n!)^{1/n}$ does not depend on the fading, $\misr_n$ approaches a non-trivial
limit as $m\to\infty$.

\begin{figure}
\subfigure[$\misr_n$ as a function of $\alpha$ for $n\in\{1,2,5\}$.]
{\epsfig{file=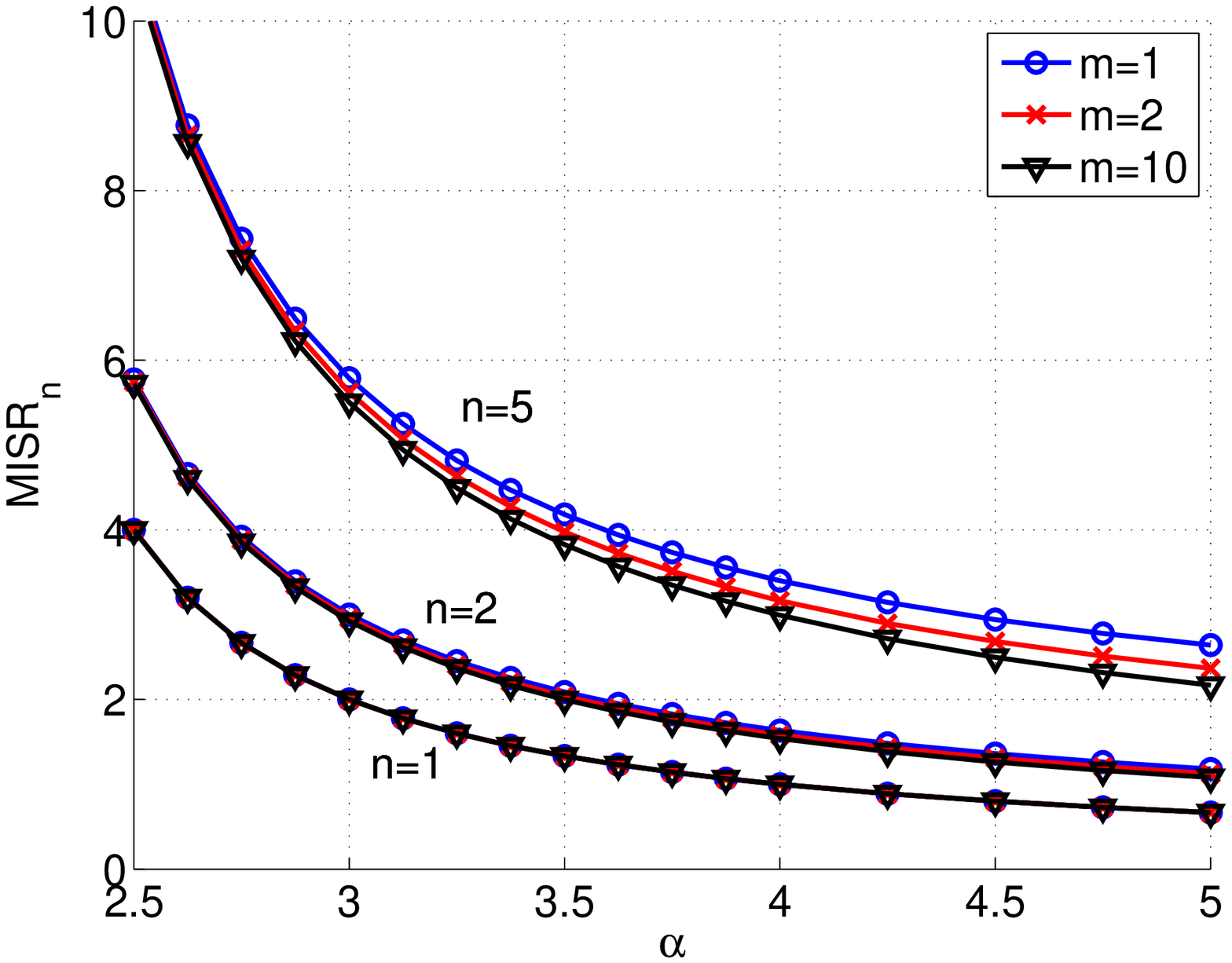,width=\figwidth}}\hfill
\subfigure[$\misr_n$ as a function of $n$ for $\alpha=4$ and lower bound \eqref{asym_delta_inf}.]
{\epsfig{file=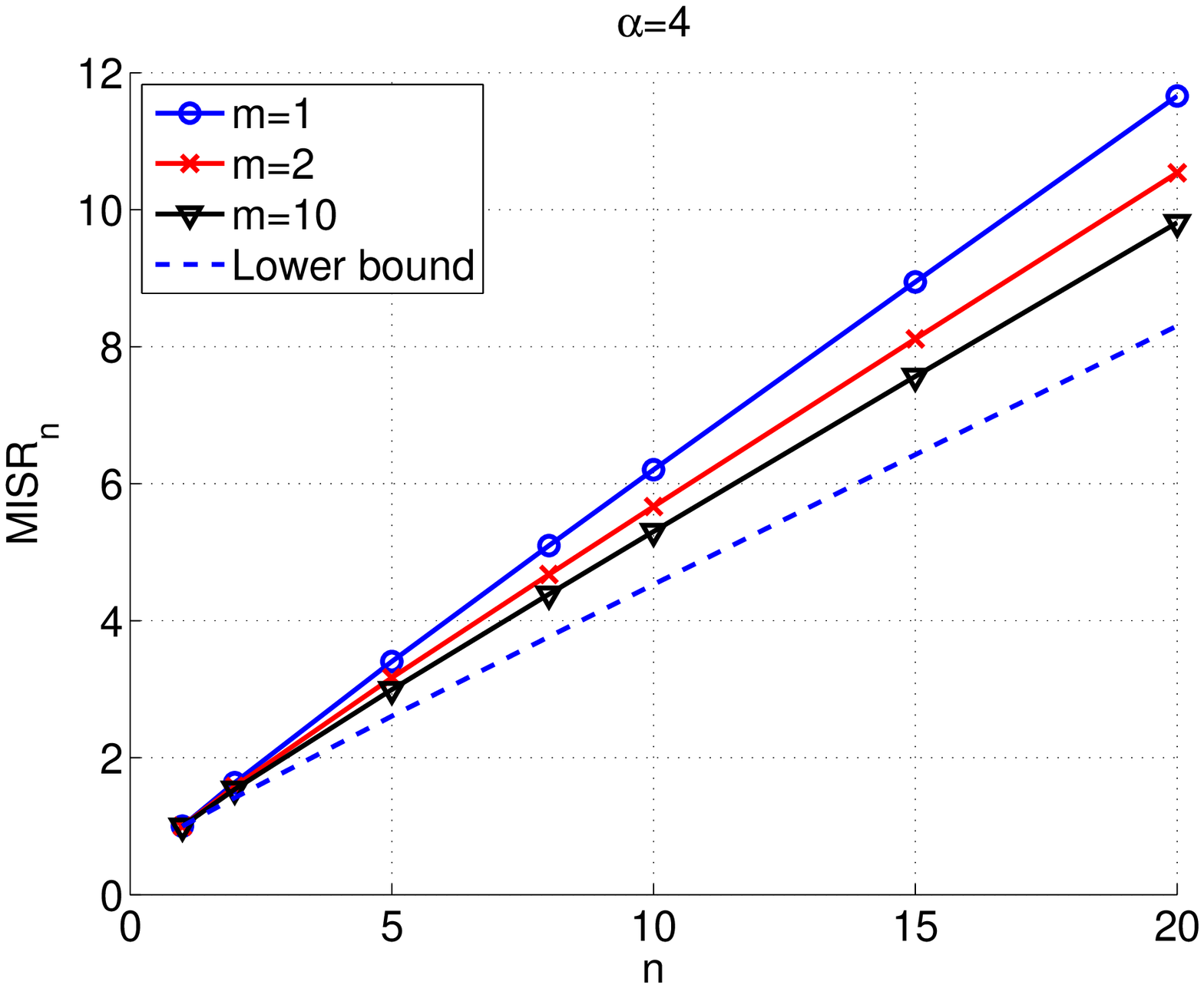,width=\figwidth}}
\caption{Generalized MISR per \eqref{gen_misr_ppp} for the PPP for Nakagami-$m$ fading with $m\in\{1,2,10\}$.}
\label{fig:gen_misr_ppp}
\end{figure}

\figref{fig:gen_misr_ppp} shows $\misr_n$ as a function of $n$. The increase is almost linear in $n$.
Indeed, as $n\to\infty$, $\misr_n$ is proportional to $n$ for the usually encountered path loss
exponents, as the following corollary establishes.

\begin{corollary}
For the PPP with Rayleigh fading and $\alpha\leq 4$,
\[ \misr_n \sim  \frac{n}{e}\,\misr_1=\frac ne \frac{\delta}{1-\delta},\quad n\to\infty. \]
\end{corollary}
\begin{IEEEproof}
For the PPP with Rayleigh fading and $\delta\geq 1/2$, it follows from  \eqref{gen_misr_ppp} that
\[ \misr_n \sim \left(\frac{\delta}{1-\delta}\right)(n!)^{1/n} ,\quad n\to\infty, \]
since the dominant term in \eqref{gen_misr_ppp} for large $n$ is the one with $\delta^n/(1-\delta)^n$, which increases
geometrically (or stays constant) with $n$ for $\delta\geq 1/2$.
For the factorial term, $\log((n!)^{1/n}) \sim \log n-1$, hence we obtain $\misr_n \sim e^{\log n-1} \,\misr_1$.
\end{IEEEproof}
{\em Remark.} Using Stirling's formula $n!\sim \sqrt{2\pi n}(n/e)^n$, this asymptotic result can be 
sharpened slightly.

\subsection{The gain $G_0$ for general fading}
Equipped with the results from Theorem \ref{thm:gen_misr_ppp}, we can now discuss the gain $G_0$ for
general fading\footnote{By ``general fading", here we refer to a fading distribution that satisfies
$F_h(x)=\Theta(x^m)$, $x\to 0$, for arbitrary $m\in\N$.}.
If $F_h(x)\sim c_m x^m$, $x\to 0$, then, for $\theta\to 0$, we have $\ps(\theta) \sim 1-c_m\, \E[(\theta\,\isr)^m]$, hence
\begin{equation}
  G_0^{(m)}=\left(\frac{\E(\isr_{\rm PPP}^m)}{\E(\isr^m)}\right)^{1/m}
 =\frac{\misr_{m,{\rm PPP}}}{\misr_m}.
 \label{g0_a}
\end{equation}
The ASAPPP approximation follows as
\[ \ps(\theta) \approx \psppp^{(m)}(\theta/G_0^{(m)}), \]
where $\psppp^{(m)}$ is the success probability for the PPP with fading parameter $m$, which is not known in closed-form. In  \cite{net:Ganti13icc}, the SIR ccdf for a Poisson cellular network  when $h$ is gamma distributed is discussed.
 However, we have the exact $\misr_m$ from \eqref{gen_misr_ppp} and the lower bound
$\misr_m\gtrsim \misr_1 (m!)^{1/m}$. 

 \begin{figure}
\centerline{\includegraphics[width=\figwidth]{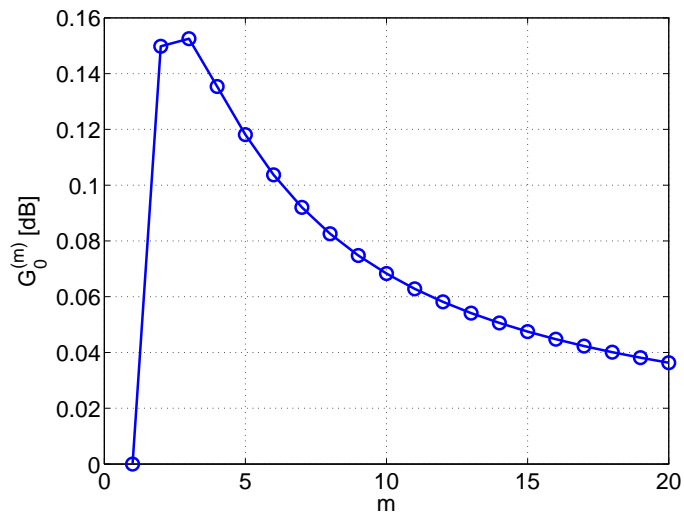}}
\caption{Gain of selection combining of $m$ transmissions in Rayleigh fading channels over
a single transmission in Nakagami-$m$ fading channels, for Poisson networks with $\alpha=4$.}
\label{fig:selcomb_vs_naka}
\end{figure}

For Nakagami-$m$ fading, the pre-constant is $c_m=m^{m-1}/\Gamma(m)$,
and we have
\begin{align*}
 \psppp^{(m)}(\theta) & \sim 1-c_m\E[(\theta\,\isr)^m] \\ 
  & \lesssim 1-\frac{m^{m-1}}{\Gamma(m)}\misr_1 m!\theta^m \\
  & =1-\misr_1(m\theta)^m ,
\end{align*}
where '$\lesssim$' indicates an upper bound with asymptotic equality.
Adding the second term in the lower bound and noting that
\[ \E(h^m)=\frac{\Gamma(2m)}{\Gamma(m)m^m} \]
yields the slightly sharper result
\[ \psppp^{(m)}(\theta) \lesssim 1-\theta^m\left[\left(\frac{m\delta}{1-\delta}\right)^m+
   \frac{\delta}{m-\delta}\frac{\Gamma(2m)}{\Gamma(m)m^m}\right] .\]
 
The gain for general fading is applicable to arbitrary transmission techniques that provide the same
amount of diversity, not just to compare different base station deployments.
 As an example, we determine the gain from selection combining of the signals from $m$ transmissions
 over Rayleigh fading channels with a single transmission over Nakagami-$m$ fading channels, both for
 Poisson distributed base stations.
The MISR for the selection combining scheme follows from \cite[Prop.~3]{net:Zhang14twc}. 
\figref{fig:selcomb_vs_naka} shows that there is a very small gain from selection combining.

Simulation results indicate that at least for moderate $m$, the scaling $\misr_m\approx \misr_1(m!)^{1/m}$ holds
 for arbitrary motion-invariant point processes. This implies that $G_0^{(m)}\approx G_0^{(1)}$,
which indicates that $G_0$ is insensitive to the fading statistics for small to moderate $m$.
\mh{We should try to say something a bit sharper here.}
Next we show that the gain is also insensitive to the path loss exponent $\alpha$.
\rk{We can make a claim that $\misr_m\geq \misr_1(m!)^{1/m}$ if we assume $\beta_m(t_1,t_2,\hdots,t_m)\geq  \prod_{k=1}^m\beta_1(t_k)$. Currently, I dont know if the condition is true for any PP.  What do you think? Otherwise we drop this paragraph for now.}

\subsection{Insensitivity of the MISR to $\alpha$}
\label{sec:insens}
\begin{figure}
\subfigure[Relative intensity $\lambda_{\rm sq}(r)/\lambda_{\rm PPP}(r)$ of square lattice.]{\includegraphics[width=\figwidth]{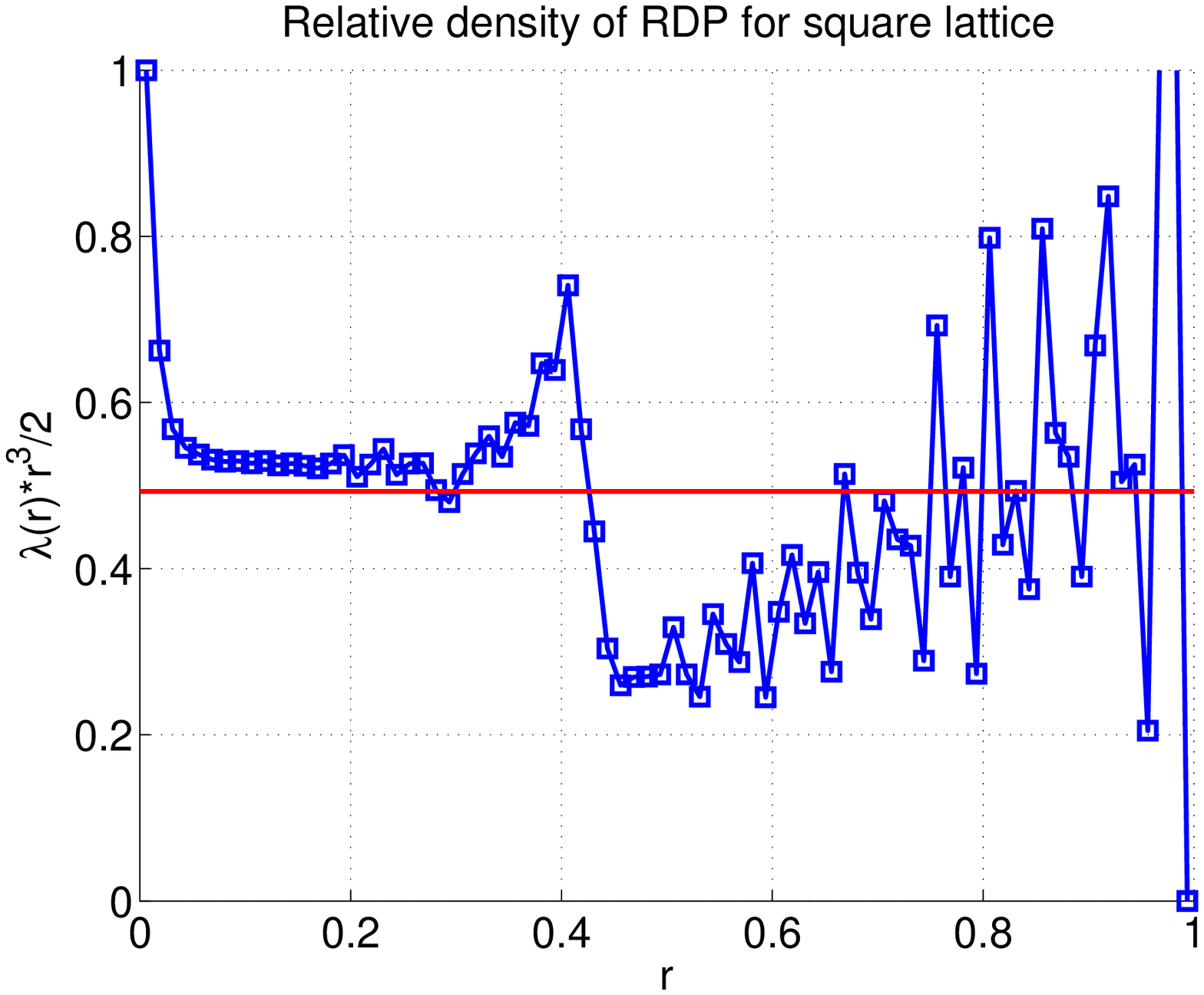}}
\subfigure[Relative intensity $\lambda_{\rm tri}(r)/\lambda_{\rm PPP}(r)$ of triangular lattice.]{\includegraphics[width=\figwidth]{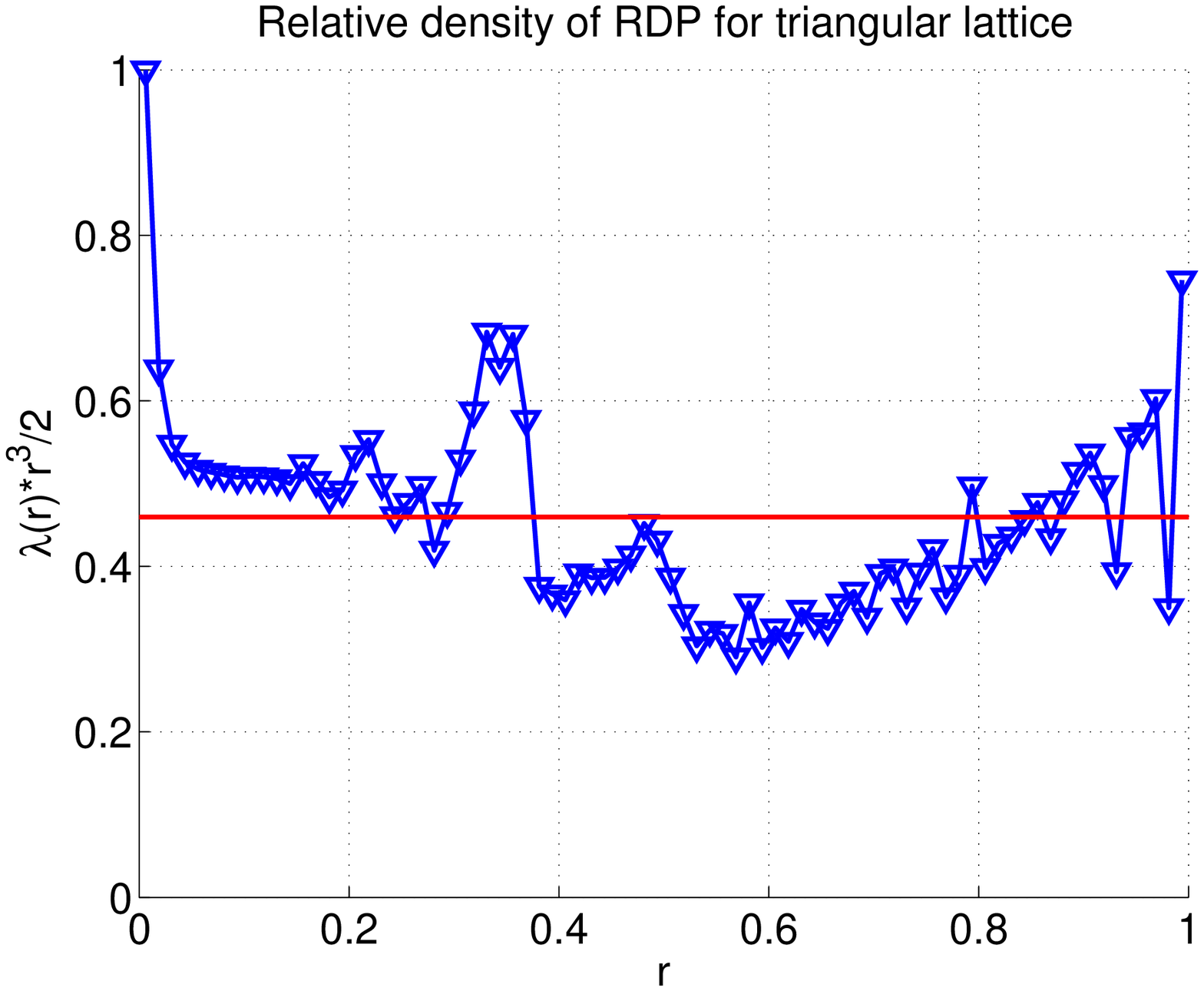}}
\caption{Relative intensity of square and triangular lattices.
The straight line is the average relative intensity and corresponds
to $1/G_{0,{\rm sq}}$ and $1/G_{0,{\rm tri}}$, respectively.}
\label{fig:rdp_density}
\end{figure} 

 \figref{fig:rdp_density} illustrates the densities of the square and triangular lattices relative to the PPP's, which is
$\lambda_{\rm PPP}(r)=2r^{-3}$, $r\in(0,1]$, as derived after Def.~\ref{def:rdp_def}.
Since the relative densities are roughly constant over the $[0,1]$ interval, the gains do not depend strongly on $\alpha$.
Indeed, if the density of the RDP of a general point process could be expressed as $\lambda(r)=c\lambda_{\rm PPP}(r)$,
we would have $G_0=1/c$ irrespective of $\alpha$.

 Another way to show the insensitivity of the gain to $\alpha$ is by exploring the asymptotic behavior
 of the MISR for general point processes given in Theorem \ref{thm:misr_pp} in the high-$\alpha$ regime.
 The result is the content of the next lemma.
  
 \begin{lemma}
For a motion-invariant point process $\Phi$,  
\begin{align}
\misr(\delta) \sim\delta \beta_1(1),\quad \delta\to 0,
\end{align}
where 
 \begin{equation*}
\beta_1(1)=\frac{1}{2}\int_{\R^2}  \|y_0\|^{2} 
 \int\limits_{[0, 2\pi]} \E^{!}_{y_0,\left(\|y_0\|,\varphi_1\right) }[g(\Phi,y_0)]\rho^{(2)}_\Phi\left(y_0,\left(\|y_0\|,\varphi_1\right)\right) \dd \varphi_1  \dd y_0.
 \end{equation*}
\end{lemma}
\begin{IEEEproof}
The $\misr$ for a general point process is given by Theorem \ref{thm:misr_pp} as
\begin{equation}
  \misr =  2\int_0^1 t^{\alpha -3}\beta_1(t)\dd t = 2 \int_0^1 t^{-3} e^{\alpha \log(t)} \beta_1(t)\dd t.
  \label{eq:misr_int}
\end{equation}
Using the Laplace asymptotic technique \cite[Eq. 6.419]{net:Orszag78book}, 
\begin{equation*}
\misr(\alpha) \sim 2  \alpha^{-1} \beta_1(1), \quad  \alpha \to \infty.
\end{equation*}
\end{IEEEproof}
This shows that $\misr$ for arbitrary point processes decays as $1/\alpha$,
which implies $G_0$ approaches a constant for large $\alpha$ (see \figref{fig:gen_misr_ppp}(a)). 

\mh{Actually, maybe this result deserves to be stated not just in the context of the sensitivity to $\alpha$.
The asymptotics $G_0\sim (\beta_1(1)-\delta \beta'_1(1)/2)^{-1}$ are interesting on their own right.}
\rk{ That is true. I think we will push the result to some other paper for a later stage. We can get more refined results (As you have pointed in your email).}

\section{The Expected Fading-to-Interference ratio (EFIR) and the Gain at $\infty$}
In this section, we define the {\em expected fading-to-interference ratio (EFIR)} and explore
its connection to the gain $G_{\infty}$ in \eqref{g_infty}. We shall see that the EFIR plays
a similar role for $\theta\to\infty$ as the MISR does for $\theta\to 0$.

\subsection{Definition and EFIR for PPP}
\begin{definition}[Expected fading-to-interference ratio (EFIR)]
\label{def:efir}
For a point process $\Phi$, let $I_\infty=\sum_{x\in\Phi} h_x \|x\|^{-\alpha}$ and let
$h$ be a fading random variable independent of all $(h_x)$.
The {\em expected fading-to-interference ratio} (EFIR) is defined as
\begin{equation}
\efir \triangleq  %\left(\lambda\pi\E^{!o} (I_\infty^{-\delta})\E(h^\delta))\right)^{1/\delta}
\left(\lambda\pi\E^{!}_o \left[\left(\frac{h}{I_\infty}\right)^\delta\right]\right)^{1/\delta} ,
\label{efir_def}
\end{equation}
where $\E^{!}_o$ is the expectation with respect to the reduced Palm measure of $\Phi$.
\end{definition}
Here we use $I_\infty$ for the interference term, since the interference here is the total received power from all points in $\Phi$,
in contrast to the interference $I$, which stems from $\Phi^!$.

{\em Remark.} For the PPP, the EFIR does not depend on $\lambda$, since $\E^{!}_o(I_\infty^{-\delta})\propto 1/\lambda$.
To see this, let $\Phi'\triangleq c\Phi$ be a scaled version of $\Phi$. Then
\[ I_c\triangleq \sum_{x\in\Phi'} h_x\|x\|^{-\alpha} =  c^{-\alpha}\sum_{x\in\Phi} h_x\|x\|^{-\alpha}\]
and thus $I_c^{-\delta}=c^2 I^{-\delta}$. Multiplying by the intensities, 
$\lambda_c I_c^{-\delta}=\lambda I^{-\delta}$ since $\lambda/\lambda_c=c^2$.
The same argument applies to all point processes for which changing the intensity by a factor $c^{-2}$ is equivalent
in distribution to scaling the process by $c$, \ie, for point processes $\Phi(\lambda)$ where
$c\Phi(1)\stackrel{{\rm d}}{=}\Phi(c^{-2})$.
This excludes hard-core processes with fixed hard-core distance but includes lattices and hard-core
processes whose hard-core distance scales with $\lambda^{-1/2}$.
\begin{lemma}[EFIR for the PPP]
\label{lem:efir_ppp}
For the PPP, with arbitrary fading,
\begin{equation}
\efir_{\rm PPP}=(\sinc\delta)^{1/\delta}.
\label{efir_ppp}
\end{equation}
\end{lemma}
\begin{IEEEproof}
The term $\E^{!}_o (I_\infty^{-\delta})$ in \eqref{efir_def} can be calculated by taking the expectation 
of  the following identity which follows from the definition of the gamma function $\Gamma(x)$.
\begin{align*}
I_\infty^{-\delta} \equiv \frac{1}{\Gamma(\delta)}\int_0^\infty e^{-s I_\infty}s^{-1+\delta}\dd s. 
\end{align*}
Hence 
\begin{align}
 \E^{!}_o(I_\infty^{-\delta}) = \frac{1}{\Gamma(\delta)}\int_0^\infty \L^{!}_{o,I_\infty}(s)  s^{-1+\delta}\dd s.
\label{eq:lap1}
\end{align}
From Slivnyak's theorem \cite[Thm.~8.10]{net:Haenggi12book}, $\E^{!}_o\equiv \E$ for the PPP,
so we can replace $\L^{!}_{o,I_\infty}(s)$ by the unconditioned Laplace transform $\L_{I_\infty}(s)$,
which is well known for the PPP and
given by  \cite{net:Haenggi08now}
\[\L_{I_\infty}(s)= \exp(-\lambda\pi \E(h^\delta)\Gamma(1-\delta)s^\delta).\]
From \eqref{eq:lap1}, we have 
\begin{align*}
 \E(I_\infty^{-\delta})&= \frac{1}{\Gamma(\delta)}\int_0^\infty e^{-\lambda \pi \E[h^\delta]\Gamma(1-\delta)s^\delta} s^{-1+\delta}\dd s \\
&= \frac{1}{\lambda \pi \E(h^\delta)\Gamma(1-\delta)\Gamma(1+\delta)}=\frac{\sinc\delta}{\lambda\pi \E(h^\delta)}.
\end{align*}
So $\lambda\pi\E^{!}_o(I_\infty^{-\delta}) \E(h^\delta)=\sinc\delta$, and the result follows.
\end{IEEEproof}
Remarkably, $\efir_{\rm PPP}$ only depends on the path loss exponent. 
It can be closely approximated by $\efir_{\rm PPP}\approx 1-\delta$.

\subsection{The tail of the SIR distribution}
Next we use the representation in \eqref{eq:rep}
to analyze the tail asymptotics of the ccdf $\bar F_{\sir}$ of the SIR (or, equivalently, the success probability $\ps$).
\begin{theorem}
\label{thm:main}
For all simple stationary BS point processes $\Phi$, where the typical user is served by the
nearest BS,
\[ \ps(\theta) \sim   \left(\frac{\theta}{\efir}\right)^{-\delta}, \quad \T \to \infty.\]
\end{theorem}
\begin{IEEEproof}
From \eqref{eq:sir}, we have $\ps(\T) =    \E \bar F_h(\theta R^\alpha I) $.
Using the representation given in \eqref{eq:rep}, it follows from the Campbell-Mecke theorem that the success probability equals
\begin{multline*}
   \E\sum_{x\in\Phi} \bar F_h\left(\theta \|x\|^\alpha 
      \sum_{y\in\Phi\setminus\{x\}} h_y\|y\|^{-\alpha} \right)\one\big(\Phi(b(o,\|x\|))=0\big) \\
  =\lambda\int\limits_{\R^2} \E_o^!\bigg[ \bar F_h\Big(\theta \|x\|^\alpha \sum_{y\in\Phi_x} \! h_y\|y\|^{-\alpha}\Big) \one(b(o,\|x\|)\;\text{empty})\bigg]\dd x,
\end{multline*}
where $\Phi_x\triangleq\{y\in\Phi\colon y+x\}$
is a translated version of $\Phi$.
Substituting $x\theta^{\delta/2} \mapsto  x$,
\begin{align}
    \ps(\theta)&=\lambda\theta^{-\delta} \int_{\R^2} \E^{!}_o \bigg[\bar F_h\Big( \|x\|^\alpha \sum_{y\in\Phi_{x\theta^{-\delta/2}}} h_y\|y\|^{-\alpha}\Big)  \one(b(o,\|x\|\theta^{-\delta/2})\text{ empty})\bigg]\dd x\nonumber\\
 &\stackrel{(a)}{\sim}\lambda\theta^{-\delta} \int_{\R^2} \E^{!}_o \bar F\left(\|x\|^\alpha I_\infty \right) \dd x,\quad \theta\to\infty \label{eq:alt}\\
 &\stackrel{(b)}{=}\lambda\theta^{-\delta}  \E^{!}_o(I_\infty^{-\delta}) \int_{\R^2} \bar F_h\left(\|x\|^\alpha  \right)  \dd x,\quad \theta\to\infty,\nonumber
\end{align}
where $(a)$ follows since $\theta^{-\delta/2} \to 0$  and hence $ \one\{b(o,\|x\|\theta^{-\delta/2})\text{ empty}\} \to 1$.
The equality in $(b)$ follows by using the substitution  $x I^{1/\alpha} \to x $. 
Changing into polar coordinates, the integral can be written as
\begin{align*}
 \int_{\R^2} \bar F_h\left(\|x\|^\alpha  \right)  \dd x =& \pi\delta \int_0^\infty r^{\delta -1} \bar F_h(r)\dd r\stackrel{(a)}{=}\pi \E(h^{\delta}),
\end{align*}
where $(a)$ follows since $h\geq 0$ \cite{net:Folland13}. % is a positive random variable .
Since $\E(h)=1$ and $\delta<1$, it follows that $\E(h^\delta)<\infty$.
\end{IEEEproof}
\arxiv
For Rayleigh fading, from the definition of the success probability and Theorem \ref{thm:main},
 \[ \ps(\theta) =  \L_{\isr }(\theta)
 \sim \left(\frac{\theta}{\efir}\right)^{-\delta}, \quad \T \to \infty. \]
 Hence the Laplace transform of $\isr$  behaves as $\Theta(\theta^{-\delta})$ for large $\theta$. Hence using the Tauberian theorem in  \cite[page 445]{net:Feller70}, %{feller2008introduction}, 
 we  can infer that
  \begin{align}
  \P(\isr < x) \sim x^\delta \frac{\efir^\delta}{\Gamma(1+\delta)},  \quad x \to 0.
  \end{align}
\mh{Interesting indeed---but what is the insight we gain from this?} 
\rk{At some stage, we were thinking about the distribution of $\isr$ at the origin. Intuitively, we can get this from using the  distribution of $\sir$ at large $\theta$ by using the distribution of $1/\sir$ . However, the fading  of the received signal is not averaged out.  We can push this to the ArXiv version.}
\fi
  
From  Theorem \ref{thm:main}, the gain $G_\infty$ immediately follows.
\begin{corollary}[Asymptotic gain at $\theta\to\infty$]
For an arbitrary simple stationary point process $\Phi$ with EFIR given in Def.~\ref{def:efir}, the asymptotic gain
at $\theta\to\infty$ relative to the PPP is 
\[ G_\infty=\frac{\efir}{\efir_{\rm PPP}}= \left(\frac{\lambda\pi\E^{!}_o (I_\infty^{-\delta})\E(h^\delta)}{\sinc\delta}\right)^{1/\delta}. \]
\end{corollary}
\begin{IEEEproof}
From Theorem \ref{thm:main}, we have that the constant $c$ in \eqref{g_infty} is given by $c=\efir^\delta$.
$c_{\rm PPP}$ follows from Lemma \ref{lem:efir_ppp} as $c_{\rm PPP}=\efir_{\rm PPP}^\delta=\sinc\delta$.
\end{IEEEproof}

The Laplace transform of the interference in \eqref{eq:lap1} for general point processes
can be expressed as
\begin{align*}
 \L^{!}_{o,I_\infty}(s) & = \E^{!}_o\left(e^{-s \sum_{x\in \Phi} h_x \|x\|^{-\alpha}} \right)\\
&=  \E^{!}_o\prod_{x\in \Phi}\L_{h}(s \|x\|^{-\alpha})= \calG^{!}_o[\L_h({s\| \cdot \|^{-\alpha}})],
\end{align*}
where $\calG^{!}_o$ is the probability generating functional with respect to the reduced Palm measure and
$\L_h$ is the Laplace transform of the fading distribution.

\begin{corollary} [Rayleigh fading]
\label{cor:ray}
With  Rayleigh fading, the expected fading-to-interference ratio simplifies to
\[\efir=  \left(\lambda \int_{\R^2} \calG^{!}_o [\Delta(x,\cdot)]\dd x\right)^{1/\delta},\]
where 
\[\Delta(x,y) = \frac{1 }{1+  \|x\|^{\alpha}\|y\|^{-\alpha}}.\]
\end{corollary}
\begin{IEEEproof}
With Rayleigh fading, the power fading coefficients are exponential, \ie, $\bar F_h(x)= \exp(-x)$. From \eqref{eq:alt}, we have
\begin{align*}
\ps(\T) &\sim  \lambda \theta^{-\delta} \int_{\R^2} \E^{!}_o \bar F\left(\|x\|^\alpha I \right) \dd x\\
&= \lambda \theta^{-\delta}  \int_{\R^2} \E^{!}_o\prod_{y\in \Phi}\frac{1}{1+\|x\|^\alpha \|y\|^{-\alpha}}\dd x,
\end{align*}
and the result follows from the definition of the reduced probability generating functional. 
\end{IEEEproof}
For Rayleigh fading, the fact that $\theta^\delta \ps(\theta)\to \sinc\delta$ as $\theta\to\infty$
was derived in \cite[Thm.~2]{net:Miyoshi14aap}.

\subsection{Tail of received signal strength}
While Theorem \ref{thm:main} shows that  $\ps(\theta) = \Theta(\theta^{-\delta}), \theta \to \infty$, it is not clear, if the scaling is mainly contributed by the received signal strength or the interference. Intuitively, since an infinite network is considered, the event of the
interference being small is negligible and hence for large $\theta$, the event $S/I > \theta$ is mainly determined by the random variable $S$.
This is in fact true as is shown in the next lemma.
\begin{lemma}
For all stationary point processes and arbitrary fading, the tail of the ccdf of the
desired signal strength $S$ is
\[ \P(S>\theta) \sim\lambda\pi \E(h^\delta)\theta^{-\delta},\quad\theta\to \infty. \]
\end{lemma}
\begin{IEEEproof}
The cdf of the distance $R$ to the nearest BS is $F_R(x)\sim\lambda\pi x^2$ for all
stationary point processes \cite{net:Stoyan95}. Hence 
\begin{align*}
\P(S>\theta)&= \P(R < (h/\T)^{\delta/2})\\
&\sim \lambda\pi\E[(h/\T)^{\delta}].
\end{align*}
\end{IEEEproof}
So the tail of the received signal power $S$ is of the same order $\Theta(\theta^{-\delta})$,
and the interference and the fading only affect the pre-constant.
In the Poisson case with Rayleigh fading,
\[ \ps(\theta)\sim\lambda\pi\Gamma(1+\delta) \theta^{-\delta},\quad\theta\to \infty. \]
The same holds near $\theta=0$. If for the fading cdf, $F_h(x)\sim a x^m$, $x\to 0$,
\[ \P(S<\theta)= \E F_h(\theta R^\alpha)\sim a\theta^m \E(R^{m\alpha}) ,\quad\theta\to 0.\]
For the PPP,
\[ \P(S<\theta)\sim \frac{\Gamma(1+m\alpha/2)}{(\lambda\pi)^{m\alpha/2}}\theta^m ,\quad\theta\to 0.\]
{\em So on both ends of the SIR distribution, the interference only affects the pre-constant.}

We now explore the tail of the distribution to the maximum SIR seen by the typical user for exponential $h$. 
Assume that the typical user connects to the BS  that provides the {\em instan\-taneously} strongest  SIR
(as opposed to the strongest SIR {\em on average} as before).
Also assume that $\theta >1$. Let $\sir(x)$ denote the SIR between the BS at $x$ and the user at the origin.
Then 
\begin{align*}
\P(\max_{x\in \Phi}{\SIR(x)} > \theta ) &= \E \sum_{x \in \Phi}\P(\SIR(x)>\theta)\\
&=\lambda \int_{\R^2} \P^{!}_o(\SIR(x)>\theta)\dd x\\
&=\lambda \int_{\R^2}  \calG^{!}_o\left[\frac{1}{1+ \theta(\|x\|/\|\cdot\|)^{\alpha}}\right]\dd x\\
&=\lambda\theta ^{-\delta}  \int_{\R^2} \calG^{!}_o [\Delta(x,\cdot)]\dd x.
\end{align*}

From the above we observe that (for exponential fading),
\[\ps(\theta)  \sim     \P(\max_{x\in \Phi}{\SIR(x)} > \theta ),\quad \theta \to \infty. \]
which shows that the tail with the maximum $\sir$ connectivity coincides with the nearest neighbor connectivity. 

 \mh{What exactly is the property?}
 \rk{Not able to locate it. "Essentially, the sum is identical to the maximum and is identical to the tail of the original RV with a scaling factor." Anyway, I modified the above sentence. Essentially, it shows that for large theta, the signal fades do not matter.  }

 \section{Examples}
 \subsection{Lattices}

Let $u_1, u_2$ be iid uniform random variables in $[0,1]$. The unit intensity (square) lattice  point process $\Phi$ is
defined as 
$\Phi \triangleq  \mathbb{Z}^2+(u_1,u_2)$.  For this lattice, with Rayleigh fading,  the Laplace transform  of the interference is bounded
as \cite{net:Giacomelli11ton} 
\begin{align}
e^{-s Z(2/\delta)}\leq  \L^{!}_{o,I_\infty}(s) \leq \frac{1}{1+s Z(2/\delta)},
\end{align}
where $Z(x) =4 \zeta(x/2)\beta(x/2)$ is the  Epstein zeta function, $\zeta(x)$ is the Riemann zeta function, and
$\beta(x)$ is the Dirichlet beta function. Hence from \eqref{eq:lap1}
\[Z(2/\delta)^{-\delta} \leq \E^{!}_o(I_\infty^{-\delta})\leq \frac{\pi \csc(\pi\delta)}{\Gamma(\delta)Z(2/\delta)^\delta}.\]
The upper bound equals $(Z(2/\delta)^\delta\Gamma(1+\delta)\sinc\delta)^{-1}$, and 
it follows that for Rayleigh fading, 
\begin{align}
\frac{(\pi\Gamma(1+\delta))^{1/\delta}}{Z(2/\delta)} \leq\efir_{\rm lat} \leq \left(\frac{\pi }{\sinc\delta}\right)^{1/\delta} \frac{1}{Z(2/\delta)}.
\label{eq:lat_bou}
\end{align}
\begin{figure}
\centering
\includegraphics[width=\figwidth]{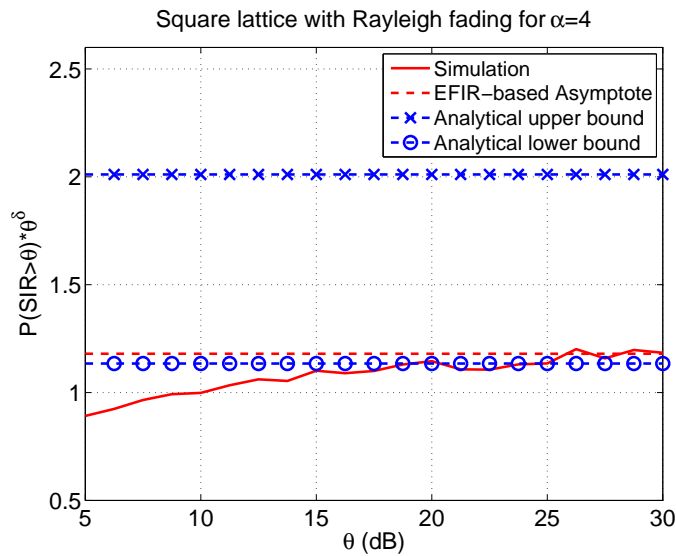}
\caption{Scaled success probability $\ps(\T)\theta^\delta$ for the square lattice point process with Rayleigh fading and $\alpha=4$.
The asymptote (dashed line) is $\sqrt{\efir}\approx 1.19$ and is tight as $\theta \to \infty$.}
\label{fig:lattice}
\end{figure} 
As $\alpha$ increases ($\delta\to 0$), the upper and lower bounds approach each other and thus both bounds get tight.

The success probability multiplied by $\theta^\delta$, the $\efir$ asymptote and its bounds \eqref{eq:lat_bou} for a square lattice process are plotted in Figure \ref{fig:lattice} for $\alpha=4$. We observe that the lower bound, which is 1.29, is indeed a good
approximation to the numerically obtained value $\efir\approx 1.40$,
and that for $\theta>15$ dB, the ccdf is already quite close to the asymptote.

\begin{figure}
\subfigure[$\alpha=3$]{\includegraphics[width=\figwidth]{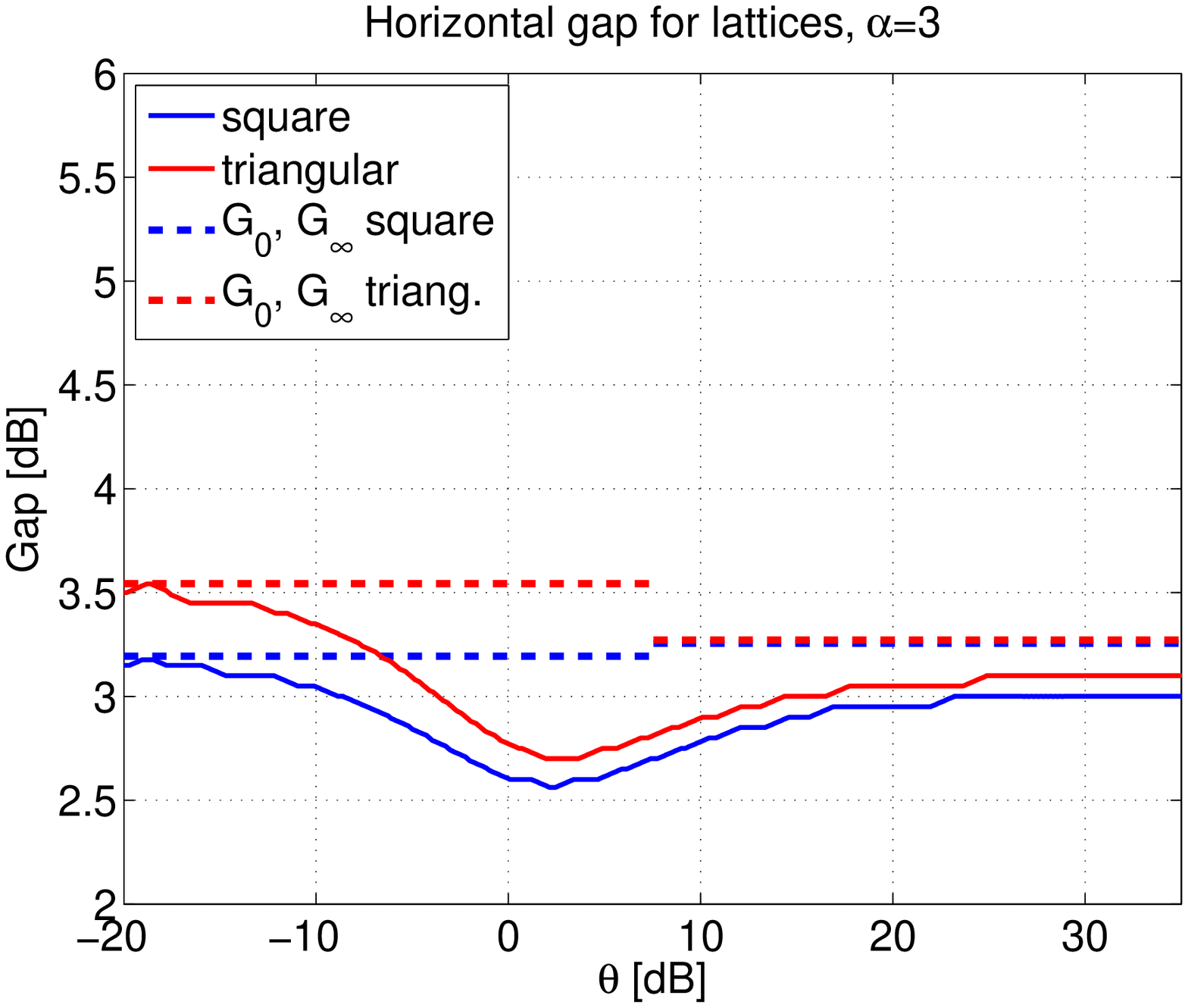}}
\subfigure[$\alpha=4$]{\includegraphics[width=\figwidth]{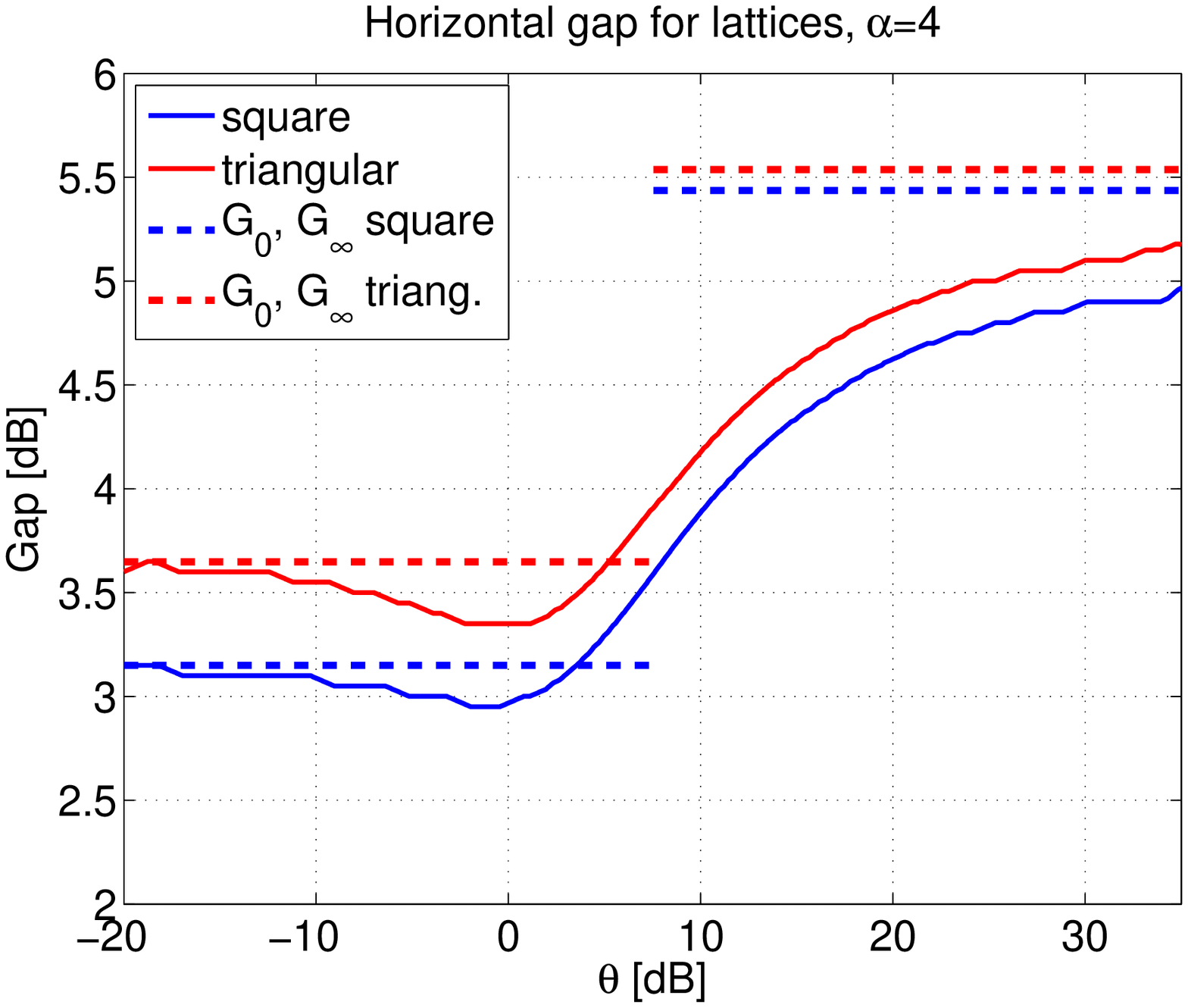}}
\caption{Gains $G(\theta)$ for the square and triangular lattices and asymptotic gains $G_0$ and $G_\infty$ (dashed) for
Rayleigh fading and $\alpha=3$ and $\alpha=4$.}
\label{fig:gains_function}
\end{figure}

 For the square and triangular lattices, \figref{fig:gains_function} shows the gain as a function of $\theta$ and
 the asymptotic gains $G_0$ and $G_\infty$ for Rayleigh fading.
 Interestingly, the behavior of the gap is not monotone. It decreases first and then (re)increases to $G_\infty$.
 It appears that $G(\theta)\leq \max\{G_0,G_\infty\}$. If this holds in general, a shift by the maximum of the two
 asymptotic gains always results in an upper bound on the SIR ccdf.
 
\begin{figure}
\centerline{\includegraphics[width=\figwidth]{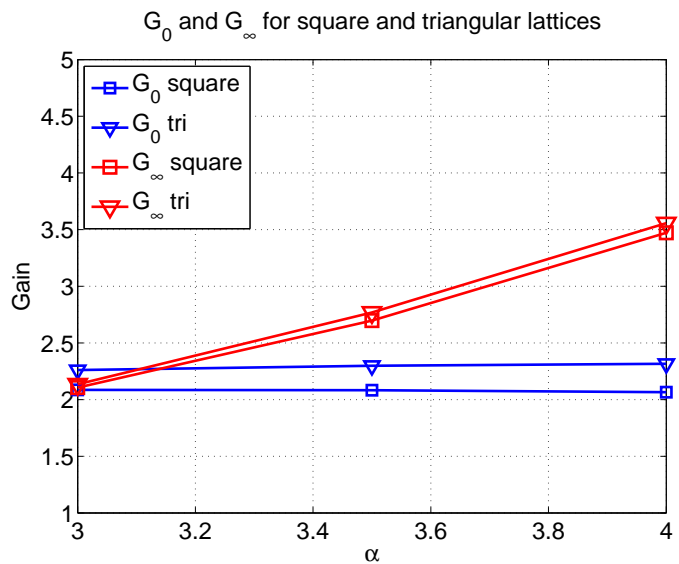}}
\caption{Asymptotic gains $G_0$ and $G_\infty$ (linear scale) for square and triangular lattices for Rayleigh fading
as a function of $\alpha$.}
\label{fig:gains_lattices}
\end{figure}

 \figref{fig:gains_lattices} shows the dependence of $G_0$ and $G_\infty$ on $\alpha$. As pointed out in
 Subs.~\ref{sec:insens}, $G_0$
 is very insensitive to $\alpha$. $G_\infty$ appears to increase slightly and linearly with $\alpha$ in this range.
 
\subsection{Determinantal point processes}
 Determinantal (fermion) point processes (DPPs) \cite{net:Hough09book} exhibit repulsion and thus
 can be used to model the fact that BSs have a minimum separation.   The kernel of the DPP $\Phi$ is denoted by $K(x,y)$ 
 and---due to stationarity---is of the form $K(x-y)$.  Its determinants yield the product densities of the DPP, hence the name.
  The reduced Palm measure $\mu^{x_o}$ pertaining
 to a DPP with kernel $K^{x_o}$ is defined as
 \begin{equation}K^{x_o}(x,y)\triangleq \frac{1}{K(x_o,x_o)}\det\left( \begin{array}{ll}
 K(x,y) & K(x,x_o) \\
 K(x_o,y) & K(x_o,x_o)
 \end{array} \right),
 \label{eq:1}
 \end{equation}
 whenever $K(x_o,x_o)>0$.
Let $K^o(x,y)$ denote the kernel associated with the  reduced Palm distribution of the DPP process. The reduced probability generating functional for a DPP is given by \cite{net:Hough09book}
\begin{align}
\calG^{!}_o[f(\cdot)] \triangleq \E^{!}_o\left[\prod_{x\in \Phi} f(x)\right]=\detF(\mathbf{1}-(1-f)K^o),
\label{eq:pgfl_dpp}
\end{align}
where $\detF$ is the Fredholm determinant and $\mathbf{1}$ is the identity operator. 
 The next lemma characterizes the EFIR a general DPP with Rayleigh fading. 
 \begin{lemma}
When the BSs are distributed as a stationary DPP, the EFIR with Rayleigh fading is 
\begin{align}
\efir =  \left(\lambda\int_{\R^2}  \detF(\mathbf{1}-(1-{\Delta}(x,\cdot))K^o)\dd x\right)^{1/\delta}.
\end{align}
 \end{lemma}
  \begin{IEEEproof}
 Follows from Corollary \ref{cor:ray} and \eqref{eq:pgfl_dpp}.
 \end{IEEEproof}
 
\begin{figure}
\centerline{\includegraphics[width=\figwidth]{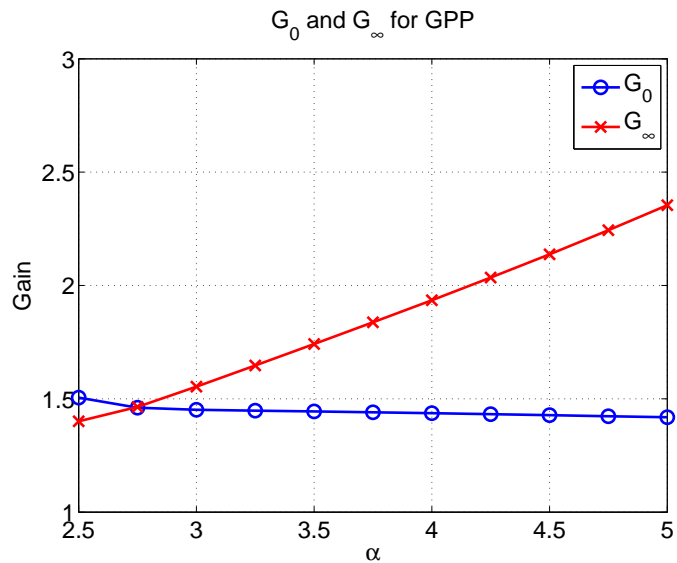}}
\caption{Simulated gains $G_0$ and $G_\infty$ for the GPP with Rayleigh fading as a function of $\alpha$.}
\label{fig:gains_gpp}
\end{figure}

{\em Ginibre  point processes: }
Ginibre point processes (GPPs) are determinantal point processes with density $\lambda =c/\pi$ and kernel
\[ K(x,y)\triangleq \frac{c}{\pi}e^{-\frac{c}{2 }(|x|^2+|y|^2)}e^{cx\bar{y}} . \]
Using the properties of GPPs \cite{net:Deng15twc},  it can be  shown that 
\[\E^{!}_o(e^{-sI_\infty}) =\prod_{k=1}^\infty \int_0^\infty \L_h(s r^{-\alpha/2})\frac{r^{k-1}e^{-cr}}{  c^{-k}\Gamma(k)}\dd r, \]
from which $\E^{!}_o(I^{-\delta})$ can  be evaluated using \eqref{eq:lap1}. 
\begin{figure}
\centering
\includegraphics[width=\figwidth]{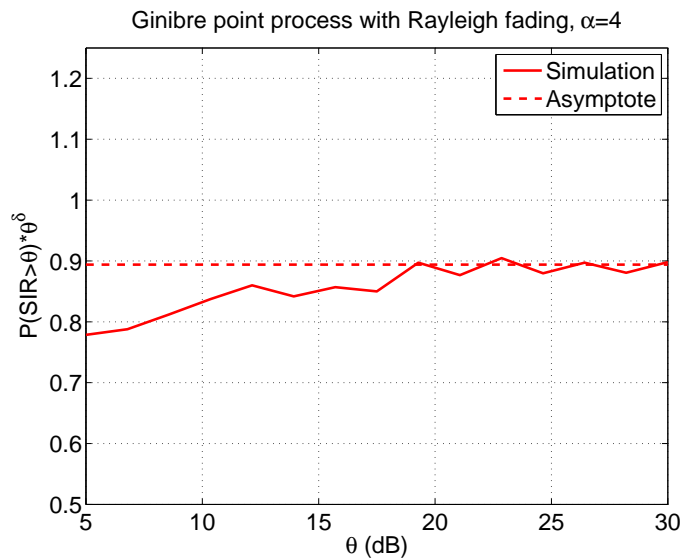}
\caption{Scaled success probability $\ps(\T)\theta^\delta$ for the GPP for Rayleigh fading with $\alpha=4$. 
The asymptote (dashed line) is at $\sqrt{\efir}\approx 0.89$.} %Computed  $\efir_{\rm DPP}\approx 0.84$.}.}
\label{fig:dpp}
\end{figure} 
In \figref{fig:dpp}, the scaled success probability $\theta^\delta\ps(\T)$ and the asymptote $\efir^\delta$ are plotted as a function of $\theta$ for the GPP. We observe a close match even for modest values of $\theta$. 
\figref{fig:gains_gpp} shows the simulated values of the gains $G_0$ and $G_\infty$ for the GPP
as a function of the path loss exponent $\alpha$. $G_0\approx 1.5$ for
all values of $\alpha$, while $G_\infty\approx \alpha/2$. 

\section{Conclusions}
This paper %analyzed the behavior of the tail of the SIR distribution for arbitrary stationary cellular network models.
established that the asymptotics of the SIR ccdf (or success probability) for arbitrary stationary cellular models
are of the form
\[ \ps(\theta)\sim 1-c_0 \theta^m,\;\;\;\theta\to 0;\;\;\quad \ps(\theta)\sim c_\infty \theta^{-\delta},\quad\theta\to\infty \]
for a fading cdf $F_h(x)=\Theta(x^m)$, $x\to 0$. Both constants $c_0$ and $c_\infty$ depend on the path loss
exponent and the point process model, and $c_0$ also depends on the fading statistics. Depending on the point
process fading {\em may} also affect $c_\infty$.
$c_0$ is related to the mean interference-to-signal-ratio (MISR). For $m=1$, $c_0=\misr$, and for $m>1$,
$c_0$ depends on the generalized MISR.
$c_\infty$ is related to the expected fading-to-interference ratio (EFIR) through $c_\infty=\efir^\delta$. 
For the PPP, $c_\infty=\sinc\delta$.
The study of the MISR is enabled by the relative distance process, which is a novel type of point process
that fully captures the SIR statistics.
A comparison of $G_0$ and $G_\infty$ shows that a horizontal shift of the SIR distribution of the PPP
by $G_0$ provides an excellent approximation of the  entire SIR distribution of an arbitrary stationary
point process.

For all the
point process models investigated so far (which were all repulsive and thus
more regular than the PPP), 
the gains relative to the PPP are between $0$ and about $4$ dB, so the shifts are
relatively modest. Higher gains can be achieved using advanced transmission techniques,
including adaptive frequency reuse, BS cooperation, MIMO, or interference cancellation.
As long as the diversity gain of the network architecture is known and the (generalized)
MISR can be calculated (or simulated), the ASAPPP method can be applied to arbitrary
cellular architectures. Such extensions will be considered in future work. A generalization
to heterogeneous networks (HetNets) is proposed in \cite{net:Wei15globecom,net:Wei16tcom}.
The method can be expected to be applicable whenever the MISR is finite.
This excludes networks where interferers can be arbitrarily close to the receiver under
consideration while the intended transmitter is further away, such as Poisson bipolar networks.

\end{document}